\documentclass[referee]{raa}            
\usepackage{overpic}
\usepackage{makecell}
\usepackage{multirow}
\usepackage{graphicx,times}
\usepackage{natbib}
\usepackage{amssymb,amsmath}
\bibpunct{(}{)}{;}{a}{}{,}

\usepackage[pagebackref=true]{hyperref}
\hypersetup{colorlinks = true, linkcolor = green, anchorcolor = red, citecolor = blue, filecolor = red, urlcolor = red}

\begin{document}

   \title{Shape model and rotation acceleration of (1685) Toro and (85989) 1999 JD6  from optical observations}

 \volnopage{ {\bf 20XX} Vol.\ {\bf X} No. {\bf XX}, 000--000}
   \setcounter{page}{1}

   \author{Jun Tian 
   \inst{1,2}, Hai-Bin Zhao\inst{1,2,3}, Bin Li\inst{1,2}
   }

\institute{ Key Laboratory of Planetary Sciences, Purple Mountain Observatory, Chinese Academy of Sciences, Nanjing 210023, China; {\it tianjun@pmo.ac.cn,  \it meteorzh@pmo.ac.cn}\\
    \and
             School of Astronomy and Space Science, University of Science and Technology of China, Hefei 230026, China\\
	\and
            Center for Excellence in Comparative Planetology, Chinese Academy of Sciences, Heifei 230026, China \\
\vs \no 
   {\small Received 20XX Month Day; accepted 20XX Month Day}
}

\abstract{The Yarkovsky-O'Keefe-Radzievskii-Paddack (YORP) effect is a net torque caused by solar radiation directly reflected and thermally re-emitted from the surface of small asteroids and is considered to be crucial in their dynamical evolution. By long-term photometric observations of selected near-Earth asteroids, it's hoped to enlarge asteroid samples with a detected YORP effect to facilitate the development of a theoretical framework. Archived light-curve data are collected and photometric observations are made for (1685) Toro and (85989) 1999 JD6, which enables measurement of their YORP effect by inverting the light curve to fit observations from a convex shape model. For (1685) Toro, a YORP acceleration $\upsilon=(3.2\pm0.3)\times10^{-9}\ \rm{rad\cdot d^{-2}}(1\sigma\ error)$ is updated, which is consistent with previous YORP detection based on different light-curve data; for (85989) 1999 JD6, it is determined that the sidereal period is $7.667749\pm 0.000009$ h, the rotation pole direction locates is at $\lambda=232\pm 2^{\circ},\ \beta = -59\pm 1^{\circ}$, the acceleration is detected to be $\upsilon = (2.4\pm0.3)\times10^{-8}\ \rm{rad\cdot d^{-2}}(1\sigma\ error)$ and in addition to obtaining an excellent agreement between the observations and model. YORP should produce both spin-up and spin-down cases. However, including (85989) 1999 JD6, the $\rm{d}\omega/\rm{d}t$ values of eleven near-Earth asteroids are positive totally, which suggests that there is either a bias in the sample of YORP detections or a real feature needs to be explained.
}

 \authorrunning{J. Tian \& H.-B. Zhao}            
 \titlerunning{YORP effects in asteroids Toro and 1999 JD6}  
 \maketitle 
	\section{Introduction}           
	\label{sect:intro}

The Yarkovsky-O'Keefe-Radzievskii-Paddack (YORP) effect is one of the mechanisms of the long-term dynamical evolution of small asteroids in the Solar system. 
It is a net torque caused by two principal mechanisms: the anisotropic reflection of sunlight and thermal emission of an asteroid. 
The YORP effect, also called non-gravitational effect, was introduced by \citet{Rubincam2000}. 
Since then, YORP has been recognized as an important mechanism of physical and dynamical evolution. 
YORP can change the rotation rate and spin-axis obliquity and affects the distribution of rotation rates and obliquities, 
especially for asteroids in the size range from $\sim$1 m to $\sim$40 km \citep{Bottke2006}.

Asteroids can be accelerated to rotational fission \citep{jacobson2016}, mass shedding \citep{scheeres2015}, 
reshape \citep{cheng2021} and creat asteroid binaries and pairs \citep{parvec2010,  margot2015}. 
Asteroids also can be decelerated to a tumbling state. 
Lowry and Taylor have directly detected that the rotation period of asteroid (54509) YORP is decreasing continuously through the light-curve data, 
which is consistent with the prediction of the YORP theory. 
To date, the YORP effect has been detected on ten near-Earth asteroids (NEAs): 
(54509) YORP, (1862) Apollo, (1620) Geographos, (3103) Eger, (25143) Itokawa, (161989) Cacus, (101955) Bennu, (68346) 2001 KZ66, (1685) Toro, 
and (10115) 1992 SK \citep{lowry2007, kaasa2007, lowry2014, taylor2007, durech2008a, durech2008b,  durech2012,  durech2018,  nolan2019,  hergen2019, zegmott2021, durech2022}. 
Crucially, all of the detections are rotational accelerations. 
While that the YORP effect can accelerate or decelerate the rotation rate of asteroids is not only a prediction given by the YORP theory\citep{Rubincam2000}, 
but also a conclusion drawn through simulation experiments \citep{rossi2009, golu2012}. 
Although the tangential-YORP (TYORP) theory -- a component of the recoil force parallel to the surface caused by re-emission of absorbed solar light from centimeter- to decimeter-sized structures on the asteroid's surface, 
has been now proposed to explain the lack of decelerating rotation asteroids \citep{golubov2014}, 
in order to explain the real feature and further verify the YORP theory, it is still necessary to continuously enlarge the asteroid samples with a detected YORP effect.

NEAs tend to be more clearly affected by the YORP effect, as they tend to be small and close proximity to the Sun. 
Therefore, the photometric data of NEAs and the published light curves are being collected, which include, but are not limited to, 
the Light Curve Database (LCDB) and the photometric data obtained by China Near-Earth Object Survey Telescope (CNEOST), to screen for NEAs. 
In order to search for NEAs whose rotations are decelerated by YORP effect, 
special attention is also paid to those targets whose rotation periods are between 6 and 9 hours, 
as they can be observed full rotation in one night which is beneficial to obtain accurate rotation periods and maximize the ability of the light-curve inversion model to detect YORP. 

In this paper, we will present a shape model and spin-state analysis of two NEAs, (1685) Toro and (85989) 1999 JD6. 
In Section 2, observing campaign of (1685) Toro is described, and the spin state and the shape model are presented as well as the approach to detect YORP-induced rotational acceleration. 
In Section 3, the collected light-curve data, results of the sidereal period and rotational pole described, and a shape model and the value of YORP rotational acceleration by the YORP model are also given.
Section 4 provides a discussion of the results and their implications, and a summary of the main conclusion is in Section 5.


\section{(1685) Toro}
\label{sect:toro}
(1685) Toro is an Apollo-type NEA discovered by C.A.Wirtanen at Mount Hamilton on July 17,1948. It's absolute magnitude is $H=14.48\pm0.13$ mag, and the assuming slope parameter $G=0.24\pm0.11$ \citep{durech2022, Warner2009}. A tentative YORP acceleration by \citet{durech2018} was based on a data set from apparitions from 1972 to 2016, then its YORP acceleration was updated by \citet{durech2022} through adding new photometric observations for 2018, 2020, and 2021. On the basis of the photometric data from 1972 to 2016, light-curve data in 2020 is collected and continuous photometric observations of (1685) Toro in February 2021 is also carried out. With all of these light-curve datasets, its YORP value is further verified. 

\subsection{Optical light-curve}
A telescope of diameter 80 cm in Yaoan Station of China is used to carry out light curve observation of Toro. The telescope is equipped with a 4128$\times$4104 Balor 17F-12 sCMOS camera, providing a field of view (FOV) of $1.6^{\circ}\times 1.6^{\circ}$ and a pixel scale of $1.4''\times1.4''$. The Johnson R filter is chosen for the imaging observations over eight nights from 11 to 18 February 2021. Each exposure takes 60 s with cadence 22 s. The raw image data is reduced following standard procedures, including bias, dark subtraction and flat-fielding correction. Eight light curves are then obtained and are labeled with IDs 9-16 in Table \ref{tab:Table 1}. New continuous photometric observations of Toro also including previously published data at the Palmer Divide Station in 2020, the light curves are labeled with IDs 1-8 in Table \ref{tab:Table 1}. It includes (1685) Toro's distance from the Sun($r$) and Earth ($\Delta$). The solar phase angle($\alpha$) at the middle point of the observation interval is given, as well as the geocentric ecliptic longitude ($\lambda_{0}$) and ecliptic latitude ($\beta_{0}$) of the asteroid, the Universal Time (UT) ``Date'' at the beginning of the night, the apparent peak-to-peak ``Amplitude'' and ``Total'' length of the light curve in Table \ref{tab:Table 1}. The other processed light curves are obtained from the ALCDEF database \citep{warner2020} and the Database of Asteroid Models from Inversion Techniques\footnote{\label{foot:itas1}\url{https://astro.troja.mff.cuni.cz/projects/damit/}} (DAMIT) web page.

\begin{table}
    \centering
	\caption{Aspect data for new observations of (1685) Toro. Each light curve has a numerical ``ID'' listed. The data sources are given in ``Reference'' : (1)\citet{warner2020}; (2) this work. Observing facility key: 40-cm SC: 40 cm Schmidt-telescope; 80-cm, Balor 17F-12: 80 cm telescope with the Balor 17F-12 sCMOS camera.}
	\label{tab:optical light-curves of Toro}
	\resizebox{\textwidth}{!}{
	\begin{tabular}{lcccccccccc} 
		\hline
		ID & UT Date & $r$ & $\Delta$ & $\alpha$ &$\lambda$&$\beta$& Amplitude& Total &Observing& Reference\\
		     &[yyyy-mm-dd]&[AU]&[AU]&[$^{\circ}$]&[$^{\circ}$]&[$^{\circ}$]&[mag]&[h]&facility &      \\
		\hline
		1&2020-06-15&0.593&1.317&47.6&337.5&5.7&0.99&2.1&40-cm SC&1 \\
       2&2020-06-17&0.572&1.304&48.2&339.3&6.2&0.96&2.2&40-cm SC&1\\
       3&2020-06-18&0.562&1.297&48.5&340.2&6.5&0.72&2.3&40-cm SC&1\\
       4&2020-06-19&0.552&1.290&48.9&341.2&6.9&0.75&2.2&40-cm SC&1\\
       5&2020-06-21&0.532&1.277&49.6&343.2&7.5&0.80&2.3&40-cm SC&1\\
       6&2020-06-22&0.523&1.270&50.0&344.3&7.8&0.95&2.3&40-cm SC&1\\
       7&2020-06-23&0.513&1.263&50.5&345.3&8.2&0.53&2.5&40-cm SC&1\\
       8&2020-06-27&0.477&1.236&52.3&349.8&9.7&0.79&2.5&40-cm SC&1\\
       9&2021-02-11&0.938&1.588&35.4&211.0&-15.8&1.21&5.5&80-cm, Balor 17F-12&2\\
      10&2021-02-12&0.933&1.593&35.0&211.0&-16.0&1.00&4.8&80-cm, Balor 17F-12&2\\
      11&2021-02-13&0.928&1.598&34.6&211.0&-16.1&0.91&6.2&80-cm, Balor 17F-12&2\\
      12&2021-02-14&0.923&1.603&34.2&210.9&-16.3&1.39&5.8&80-cm, Balor 17F-12&2\\
      13&2021-02-15&0.918&1.608&33.8&210.9&-16.5&0.99&6.0&80-cm, Balor 17F-12&2\\
      14&2021-02-16&0.913&1.613&33.4&210.8&-16.6&1.02&5.8&80-cm, Balor 17F-12&2\\
      15&2021-02-17&0.909&1.618&33.0&210.7&-16.8&1.20&6.1&80-cm, Balor 17F-12&2\\
      16&2021-02-18&0.904&1.623&32.5&210.7&-16.9&1.08&6.3&80-cm, Balor 17F-12&2\\
 		\hline
	\end{tabular}}
	\label{tab:Table 1}
\end{table}

\subsection{Shape model and YORP rotational acceleration }

The constant torque provided by the YORP effect produces a linear change in the rotation rate. Therefore, the spin-state analysis and optimization require investigation of the precise timing of the light curve. To achieve the best match between the observed data and the model, a free parameter $\upsilon\equiv d\omega/dt$ is added to the convex inversion in the YORP model, which describes the change of rotation rate $\omega$ and is optimized during the light-curve inversion together with the shape and spin parameters. This approach is the same as what used by \citet{durech2018}. If a non-zero $\upsilon$ provides a significantly better fit than $\upsilon=0$, it is interpreted as detecting rotation acceleration or deceleration. The parameter is a linear change of the rotation rate in time $\upsilon\equiv d\omega/dt$ or a quadratic change of the rotation phase in time. The rotation phase in radians $\varphi(t)$ can be expressed for any given time as:

\begin{equation}
\varphi(t)=\varphi(T_{0})+\omega(t-T_{0})+\frac{1}{2}\upsilon(t-T_{0})^2   , 
\end{equation}
\begin{flushleft}
where:
 \item {$t$}:       the time of observation(JD), 
 \item{$T_{0}$}  :             the epoch from which the model is propagated, 
 \item{$\varphi(T_{0})$}   :     initial rotation phase in radians,
 \item{$\omega$}     :     rotation rate in $\rm{rad\cdot d^{-1}}$; $\omega\equiv2\pi/P$, $P$ is rotation period in days,
 \item{$\upsilon$}    :      the change of rotation rate in $\rm{rad\cdot d^{-2}}$; $\upsilon\equiv  d\omega/dt$ (the YORP strength).
\end{flushleft}

\begin{figure}
    \centering
	\includegraphics[width=10cm]{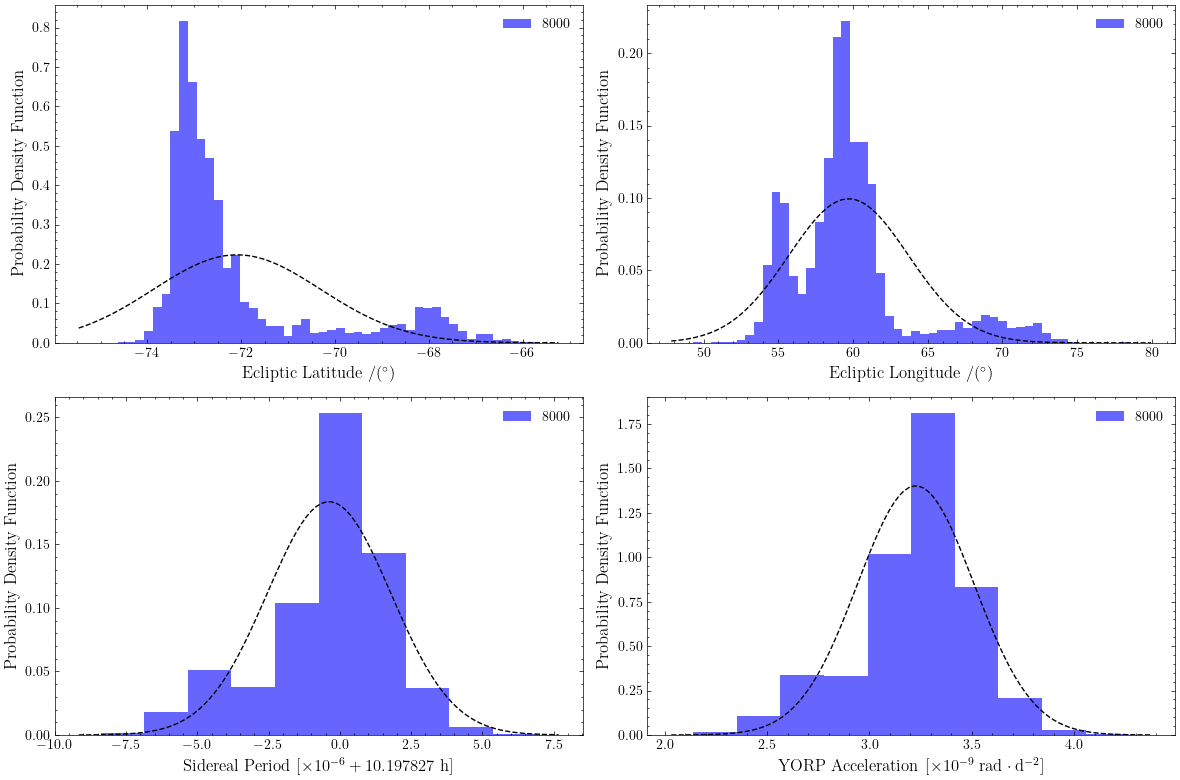}
    \caption{The uncertainties of the four parameters are calculated by the bootstrap method. The four parameters are the spin axis direction in ecliptic latitude $\beta$ and longitude $\lambda$, the sidereal period and the YORP acceleration. The black dotted line is the Gaussian fitting curve of these parameter distributions.}
    \label{fig1}
\end{figure}

The YORP model iteratively converges to best-fit parameters to minimize the difference between the observed and modeled light curves. These spin-state parameters published by \citet{durech2018} are $P=10.19782\pm0.00003$ h (for JD 2441507.00000), $(\lambda,\beta)=(71\pm10^{\circ},\ -69\pm5^{\circ})$ $(3\sigma\ \rm{errors})$, and $\upsilon=3.0\times10^{-9}\ \rm{rad\cdot d^{-2}}$.
 Our study uses these spin-state parameters as input values for further optimization of the YORP model, which is reasonable and robust as verified by independent scan and optimization calculations. Consequently,  the updated values ($1\sigma$ errors) are given as follows:$P=10.197827\pm0.000002$ h (for JD 2441507.0), $(\lambda,\beta)=(60\pm4^{\circ},-72\pm2^{\circ})$, $\upsilon=(3.2\pm0.3)\times10^{-9}\ \rm{rad\cdot d^{-2}}$, which are further optimized by the YORP model as input values. In the article uncertainties are estimated by a bootstrap method. For (1685) Toro, 98 light curves are randomly selected from all 115 light curves as input for optimizing the shape model and spin-state parameters, then the above process is repeated for 8000 times. Uncertainly are estimated from the distributions of there parameters given in Fig.\ref{fig1}. 

\begin{figure}
  \centering
	\includegraphics[width=10cm]{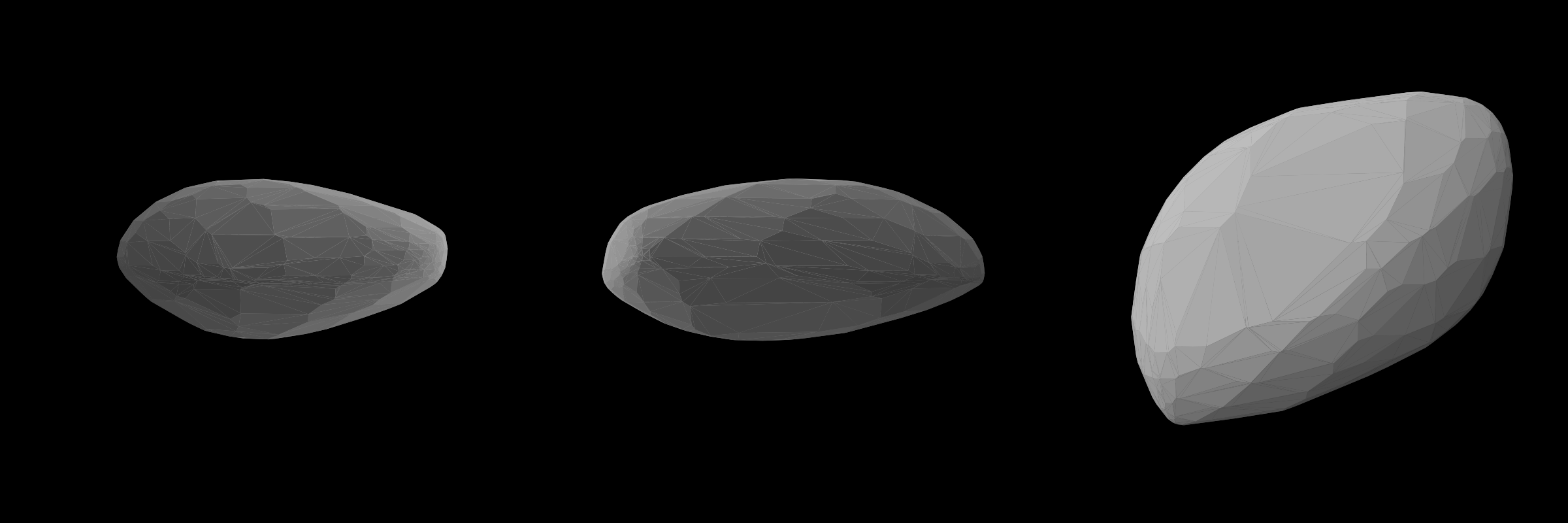}
    \caption{Shape model of asteroid (1685) Toro shown from equatorial level (y-axis, left and x-axis, center , 90$^\circ$ apart) and pole-on (z-axis, right).The model’s z-axis is aligned with the rotation axis and axis of maximum inertia.}
    \label{fig2}
\end{figure}

\begin{figure}
\centering
	\includegraphics[width=10cm]{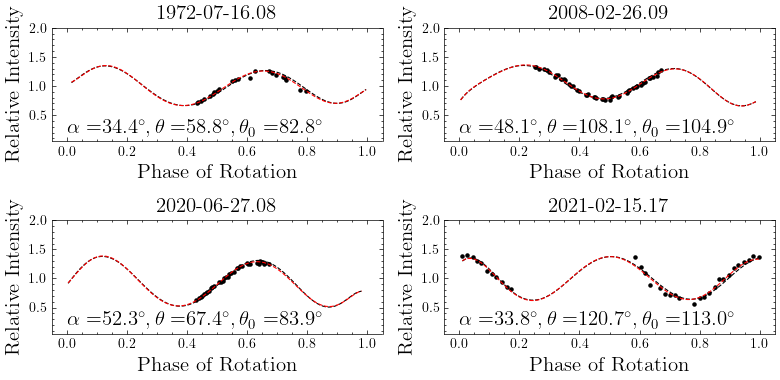}
    \caption{Example light curves (black dots) of (1685) Toro shown with the synthetic light curves produced by the best YORP model(red dashed curves) and the best constant-period model(dotted black curves). The geometry of observation is described by the aspect angle $\theta$, the solar aspect angle $\theta_{0}$, and the solar phase angle $\alpha$. A full set of light curves are provided in Fig. \ref{fig:Fig.A1}}
    \label{fig3}
\end{figure}

These spin parameters are also given by \citet{durech2022} ($1 \sigma$ error): $\upsilon=(3.3\pm0.3)\times10^{-9} \ \rm{rad\cdot d^{-2}}$ with period $P=10.197826\pm0.000002$ h (for JD 2441507.0), pole direction $(75\pm3^{\circ}, -69\pm1^{\circ})$. Except for the pole's ecliptic longitude($\lambda$) , our results are in good agreement with them. It might be a different light-curve data set used for the inversion model and different weights of individual light curves. Nonetheless, they all give a well-matched YORP rotational acceleration and sidereal period, and YORP rotational acceleration indicates a weak but robust YORP strength. The convex shape model is shown in Fig.\ref{fig2} and Fig.\ref{fig3} shows four light curves in which the difference between the two models is the largest. 

\begin{figure}
\centering
	\includegraphics[width=10cm]{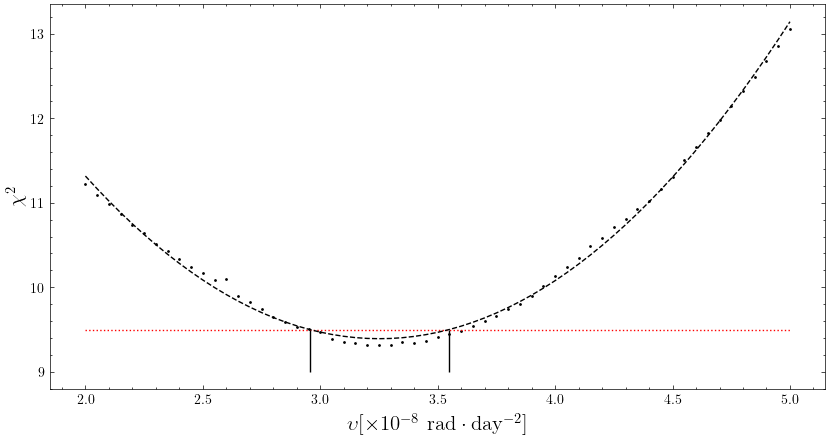}
    \caption{Dependence of the goodness of the fit measured by the $\chi^2$ on the YORP parameter $\upsilon$ for asteroid (1685) Toro. The dashed curve is a quadratic fit of the data points. The horizontal dotted line indicates a 2.0\% increase in the $\chi^2$ , which defines 1$\sigma$ uncertainty interval of $\pm0.3\times10^{-8}\ \rm{rad\cdot d^{-2}}$ given the number of degress of freedom.}
    \label{fig4}
\end{figure}

To have a realistic and independent estimate of the uncertainty of YORP acceleration, the same approach as \citet{vokh2001} and \citet{polis2014} is adopted and $\chi^2$ for different fixed values of $\upsilon$ is computed (all other parameters are optimized). Then the 1$\sigma$ uncertainty interval of  $\upsilon$ is defined so have been that $\chi^2$ increases by a factor of $1+\sqrt{2/\nu}$, where $\nu$ is the number of degrees of freedom ($\nu\ \sim$ 5000 in case of Toro). The computed $\chi^{2}$ for different fixed values of $\upsilon$ is shown in Fig. \ref{fig4}.   

Across the entire set of light-curve data, we identified a weak but reliable YORP value.  To detect the YORP effect of Toro, the observations in 1972 and 1988 are crucial. However, confirming the accuracy of the data is a challenge due to the lack of independent observations of the same apparition in 1972 and 1988. Moreover, the present light-curve data do not provide adequate constraints on the ecliptic longitude of the pole direction. Therefore, photometric observation should always be needed in the future to further confirm this YORP detection and refine the spin parameters.

\section{(85989) 1999 JD6}

(85989) 1999 JD6 is both an NEA of the Aten class and a potentially hazardous asteroid (PHA) and was discovered on May 12, 1999 at Anderson Mesa by Lowell Observatory Near-Earth-Object Search(LONEOS) project in Flagstaff \citep{mpc1999}. Photometric and radar observations indicate it is a contact-binary asteroid belonging to Barbarian asteroids. The visible spectrum data showed that it is a K-type asteroid \citep{bin2001, deleon2010}, and later it is identified as an L-type asteroid with the aid of the near-infrared spectrum. Other physical characteristics have been determined, such as albedo 0.05-0.11 \citep{campins2009, thomas2011, reddy2012, mainzer2014, nugent2016}, maximum diameter 2 km \citep{Mars2015}, rotation period $7.6638 \pm 0.0001$ h \citep{szabo2001}. Given the foregoing, is the shape model and spin-state will be given below with light-curve data.

\subsection{Optical light-curve}

The optical light-curve data for (85989) 1999 JD6 span twenty years. The first light curves of this asteroid came from the four nights of July 2,3,5, and 6, 2000, and they were observed by a 59 cm Schmidt-telescope. Light curves were published in VizieR\footnote{\label{foot:itas}\url{https://cdsarc.cds.unistra.fr/}}  \citep{szabo2001}. In addition , light-curve data also includes previously published photometry obtained at the Palmer Divide Station \citep{warner2014, warner2015, warner2018, warner2019, warner2020}. The processed light curves are retrieved from the Asteroid Light-curve Data Exchange Format (ALCDEF) database \citep{Warner2011}.

The observational circumstances of the data used in this paper are summarized in Table \ref{tab:Table 2}. It includes (85989) 1999 JD6's distance from the Sun($r$) and Earth ($\Delta$). The solar phase angle($\alpha$) at the middle point of the observation interval is given, as well as the geocentric ecliptic longitude ($\lambda_{0}$) and ecliptic latitude ($\beta_{0}$) of the asteroid, the Universal Time (UT) ``Date'' at the beginning of the night, the apparent peak-to-peak ``Amplitude'' and ``Total'' length of the light curve in Table \ref{tab:Table 2}. 

\begin{table}
    \centering
	\caption{A list of optical light curves of asteroid (85989) 1999 JD6 used in this study. Each light curve has a numerical ``ID'' listed. The data sources are given in ``Reference'': (3) \citet{szabo2001}; (4) \citet{warner2014}; (5) \citet{warner2015}; (6) \citet{warner2018}; (7) \citet{warner2019}; (8) \citet{warner2020}. Observing facility key: 59-cm ST, 59 cm Schmidt-telescope, and PDS, Palmer Divide Station(California, USA).}
	\label{tab:optical light-curves of 1999 JD6}
	\resizebox{\textwidth}{!}{
	\begin{tabular}{lcccccccccc} 
		\hline
		ID & UT Date & $r$ & $\Delta$ & $\alpha$ &$\lambda$&$\beta$& Amplitude& Total &Observing& Reference\\
		     &[yyyy-mm-dd]&[AU]&[AU]&[$^{\circ}$]&[$^{\circ}$]&[$^{\circ}$]&[mag]&[h]&facility &      \\
		\hline
		 1&2000-07-02&0.379&1.337&27.6&266.7&34.9&1.22&2.6&59-cm ST&3\\
        2&2000-07-03&0.378&1.336&27.8&266.2&34.9&1.22&2.2&59-cm ST&3\\
        3&2000-07-05&0.371&1.323&29.6&261.9&34.0&1.24&5.2&59-cm ST&3\\
        4&2000-07-06&0.370&1.321&29.8&261.4&33.9&1.24&3.6&59-cm ST&3\\
        5&2014-05-20&0.533&1.394&35.9&275.9&42.7&1.18&5.5&PDS&4\\
        6&2014-05-21&0.530&1.397&35.4&274.9&42.9&1.15&5.6&PDS&4\\
        7&2014-05-22&0.528&1.399&34.9&273.9&43.1&1.04&3.1&PDS&4\\
        8&2015-06-07&0.575&1.336&45.1&323.1&22.0&0.92&3.2&PDS&5\\
       9&2015-06-11&0.526&1.317&44.9&324.2&22.8&1.09&3.4&PDS&5\\
       10&2015-06-12&0.514&1.312&44.9&324.5&22.9&1.07&3.5&PDS&5\\
       11&2015-06-14&0.490&1.302&44.8&325.1&23.4&1.13&3.4&PDS&5\\
       12&2015-06-15&0.477&1.297&44.8&325.4&23.6&1.18&3.1&PDS&5\\
       13&2018-06-01&0.517&1.180&58.8&167.5&41.2&1.41&4.0&PDS&6\\
       14&2018-06-02&0.527&1.188&58.2&168.9&40.6&1.10&4.4&PDS&6\\
       15&2018-06-03&0.536&1.195&57.6&170.1&39.9&1.24&4.8&PDS&6\\
       16&2018-06-04&0.546&1.202&57.0&171.3&39.3&1.25&4.4&PDS&6\\
       17&2019-06-03&0.522&1.437&29.3&269.2&41.1&0.89&4.0&PDS&7\\
       18&2019-06-04&0.520&1.438&28.9&268.0&41.1&0.96&4.1&PDS&7\\
       19&2019-06-05&0.518&1.439&28.5&266.8&41.1&1.06&4.0&PDS&7\\
       20&2019-06-06&0.516&1.439&28.2&265.6&41.1&1.13&4.5&PDS&7\\
       21&2019-06-07&0.514&1.440&28.0&264.4&41.0&1.17&4.5&PDS&7\\
       22&2020-06-18&0.445&1.231&51.6&337.8&20.0&0.85&2.7&PDS&8\\
       23&2020-06-19&0.432&1.224&52.0&338.6&20.1&1.39&3.2&PDS&8\\
       24&2020-06-20&0.420&1.217&52.3&339.3&20.2&1.12&2.5&PDS&8\\
       25&2020-06-21&0.408&1.211&52.6&340.1&20.3&0.92&1.9&PDS&8\\
       26&2020-06-22&0.395&1.204&53.0&341.0&20.5&1.28&3.3&PDS&8\\
       27&2020-06-23&0.383&1.197&53.5&341.8&20.6&1.20&3.3&PDS&8\\
 		\hline
	\end{tabular}}
	\label{tab:Table 2}
\end{table}

\subsection{Sidereal period and pole search}

\begin{figure}
\centering
	\includegraphics[width=10cm]{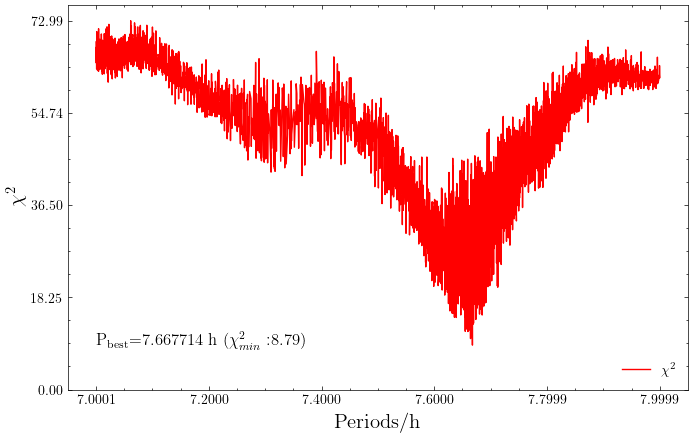}
    \caption{ This plot shows the best sidereal period (P = 7.667714 h) with the local minimum of $\chi^2$. The found best sidereal period found is used as the input value to search for the pole direction.}
    \label{fig5}
\end{figure}

The convex inversion program starts from initial values , which includes an initial period, finds the local minimum of $\chi^2$, and obtains the corresponding shape model and period solution. The method described in \citet{Kaasa2001}, a range of period values and six initial poles for each trial period are scanned across, from which the lowest $\chi^2$ value is selected. In consideration of the previously reported periods for (85989) 1999 JD6 \citep{szabo2001, warner2014, warner2015, warner2018, polis2008, warner2019, warner2020} and saving computer calculation time, scanned period value range from 7.0 to 8.0 h. The search result of the best period indicates the initial rotational period is at 7.667714 h (Fig. \ref{fig5}). This value is utilized in the next step as an optimal starting point for the subsequent optimization process.

\begin{figure}
\centering
	\includegraphics[width=10cm]{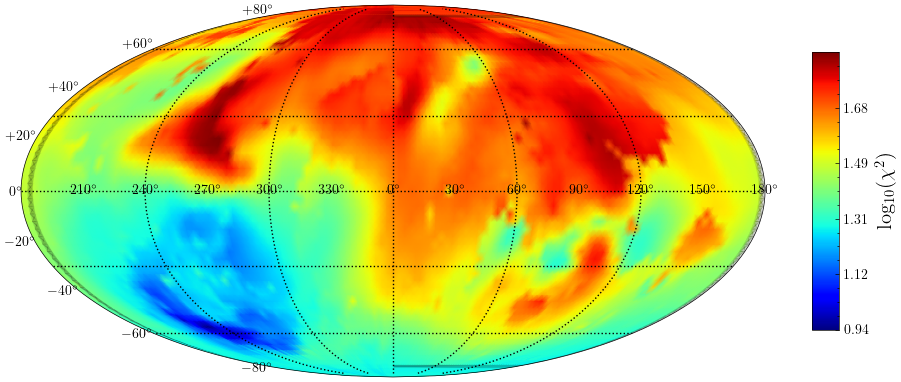}
    \caption{The $\chi^2$ values map for all possible orientations of the pole direction given in an Aitoﬀ projection of the sky in ecliptic coordinates. These results are obtained by an inversion method using only light-curve data. There is one local minima for pole position (lowest $\chi^2$ values are indicated in dark blue), and the corresponding with pole direction is $(\lambda, \beta)=(231^\circ,-60^\circ)$, where $\lambda$ and $\beta$ are pole ecliptic longitude and latitude, respectively. }
    \label{fig6}
\end{figure}

The next step in the shape modeling procedure is to search for the pole direction with a rotational period of 7.667714 h as an input value and to determine a best-fitting convex shape for 1999 JD6. The convex inversion techniques described by \citet{kaasal2001} and \citet{Kaasa2001} is applied in the shape model. The first thing to do is to set up a grid of pole positions covering the entire celestial sphere with a resolution of $3^\circ\times3^\circ$.  The $\chi^2$ is calculated to fit the light curves at each fixed pole position and the best sidereal period, and the corresponding of the minimum of $\chi^2$ value is recorded. At this stage, the initial epoch $T_{0}$ and the initial rotation phase $\varphi(T_{0})$, are needed for the transformation between vectors $\mathbf{r}_{\rm{ast}}$ in the asteroid co-rotating coordinate frame and vectors $\mathbf{r}_{\rm{ecl}}$ in the ecliptic coordinate frame, and are fixed during the optimization. So $T_{0}$ is set to be 2451728.000000, corresponding to the date of the first light curve (2000-07-02), and $\varphi(T_{0})$ is set to be $0^\circ$. This model assumes a constant rotation period and the results of the pole search are shown in Fig.\ref{fig6}. As can be seen, the pole coordinates are well constrained, thanks to the large range of geocentric ecliptic latitude sampled by the light-curve dataset.

A model with a constant rotation period provides the best pole direction $(\lambda, \beta)=(231^\circ,-60^\circ)$ and rotation period $P=7.667714$ h. However, this constant-period model is significantly worse than the YORP model in fitting the light curves. In the following subsections, these spin parameters are further optimized by the YORP  model as input values.

\subsection{Shape model and YORP rotational acceleration }

Using the YORP model, the shape model and the spin-state parameters of 1999 JD6 are obtained with the whole light-curve data and fit very well, and their uncertainties are estimated by a Monte Carlo (MC) method.  However, there are no photometric error data for four light curves in 2000. To realistically estimate uncertainties of light-curve data, a Fourier series of maximum order determined by an F-test is used to fit the light curves \citep{magnuss1996}:

\begin{equation}
F_{n}=\frac{(\chi^2_{0}-\chi^2_{n})/(\nu_{0}-\nu_{n})}{\chi^2_{n}/\nu_{n}},
\end{equation}
where $\chi^2_{n}$ is the chi-square for the fit of a Fourier series truncate after order n, with $\nu_{n}=N-(2n+1)$ degrees of freedom, and N is the number of each light curve point. $\chi^2_{0}$ is the chi-square for the light curve, with $\nu_{0} =N$ degrees of freedom. The root-mean-square residual is used as the uncertainty of individual light curve points. (85989) 1999 JD6's light curves are weighted according to their precision. 

\begin{figure}
\centering
	\includegraphics[width=10cm]{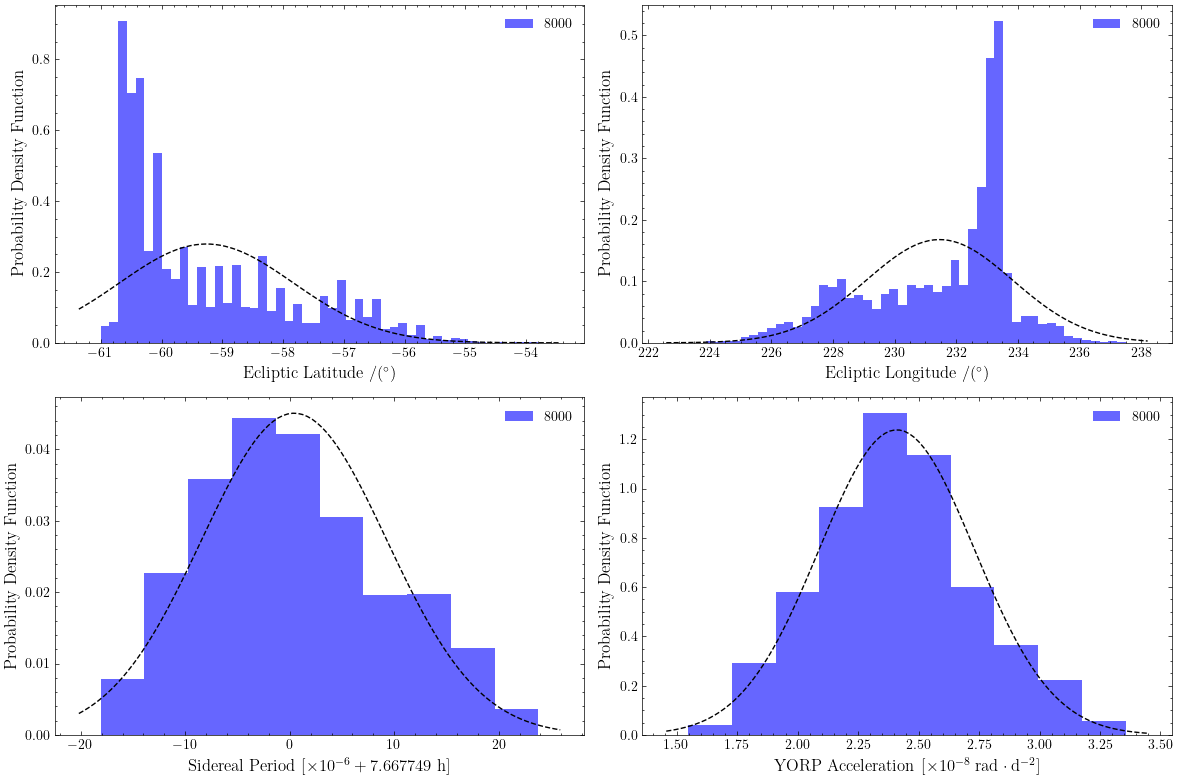}
    \caption{The uncertainties of the four parameters are calculated by the MC method. The four parameters are the spin axis direction in ecliptic latitude $\beta$ and longitude $\lambda$, the sidereal period and the YORP acceleration. The black dotted line is the Gaussian fitting curve of these parameter distributions.}
    \label{fig7}
\end{figure}

8000 virtual light-curve data sets are created by randomly adding noise within its corresponding photometric error range. Each new data set, is repeatedly inverted and spin parameters are obtained. From the distribution of these parameters in Fig.\ref{fig7}, their uncertainties are estimated. 

\begin{figure}
\centering
	\includegraphics[width=10cm]{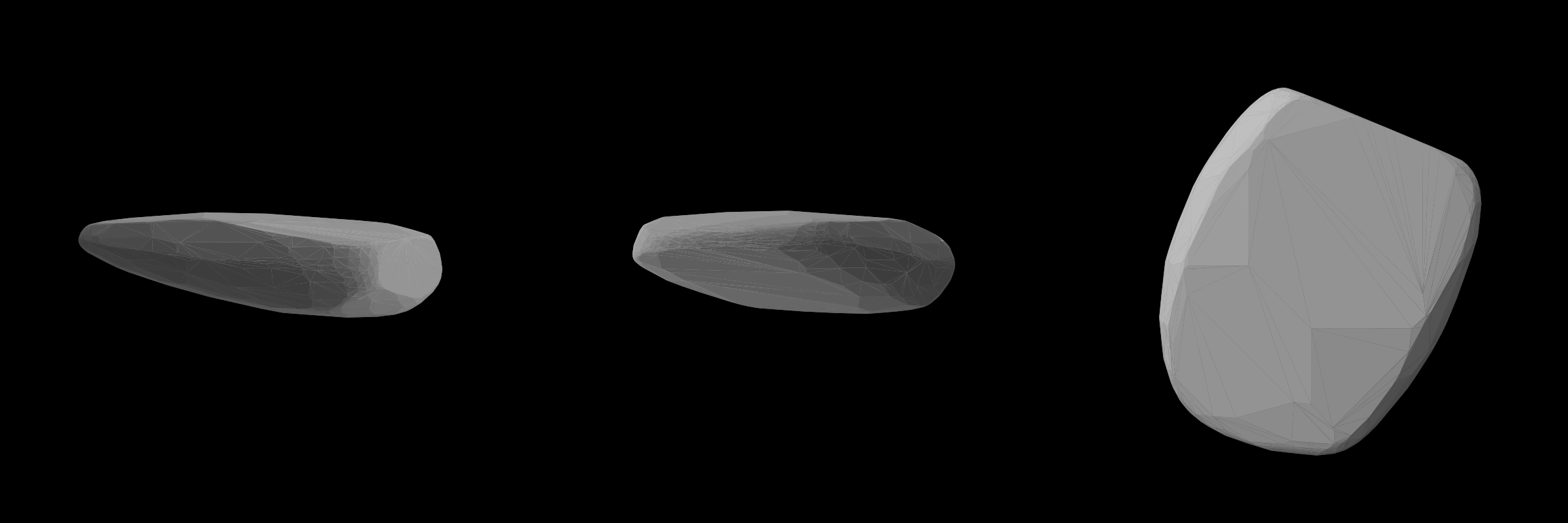}
    \caption{Shape model of asteroid (85989) 1999 JD6 shown from equatorial level (y-axis, left and x-axis, center , 90$^\circ$ apart) and pole-on (z-axis, right).The model’s z-axis is aligned with the rotation axis and axis of maximum inertia.}
    \label{fig8}
\end{figure}

\begin{figure}
\centering
	\includegraphics[width=10cm]{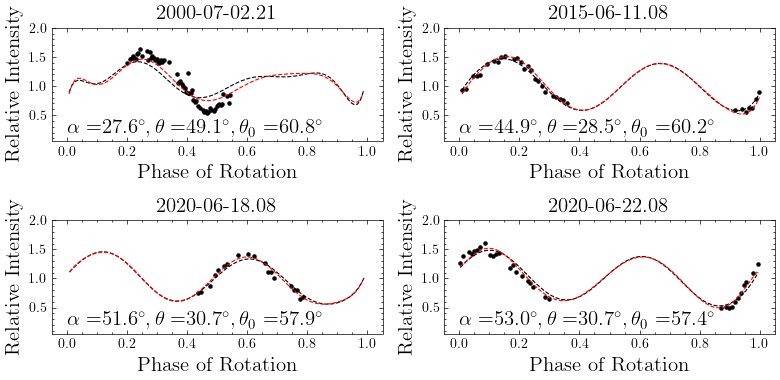}
    \caption{Example light curves (black dots) of (85989) 1999 JD6 shown with the synthetic light curves produced by the best YORP model(red dashed curves) and the best constant-period model(dotted black curves). The geometry of observation is described with the aspect angle $\theta$, the solar aspect angle $\theta_{0}$, and the solar phase angle $\alpha$. A full set of light curves are provied in Fig. \ref{fig:Fig.A2}.}
    \label{fig9}
\end{figure}

The YORP model gives the best-fit values including a pole direction in ecliptic coordinates $(232\pm2^\circ, -59\pm1^\circ)$, the sidereal period $P=7.667749\pm0.000009\ \rm{h}$ (for JD 2451728.0), and the YORP rotational acceleration $\upsilon=(2.4\pm0.3)\times10^{-8} \ \rm{rad\cdot d^{-2}}\ (1\sigma \ errors)$. The errors of these parameters are standard deviation of 1$\sigma$ uncertainties. Fig.\ref{fig8} shows the convex shape model, which is quite elongated and flat shape. The planar features are the result of the procedure attempting to match the large amplitude of the light curves. The agreement between synthetic light curves produced by this shape and real observations is demonstrated in Fig.\ref{fig9}. 

\begin{figure}
\centering
	\includegraphics[width=10cm]{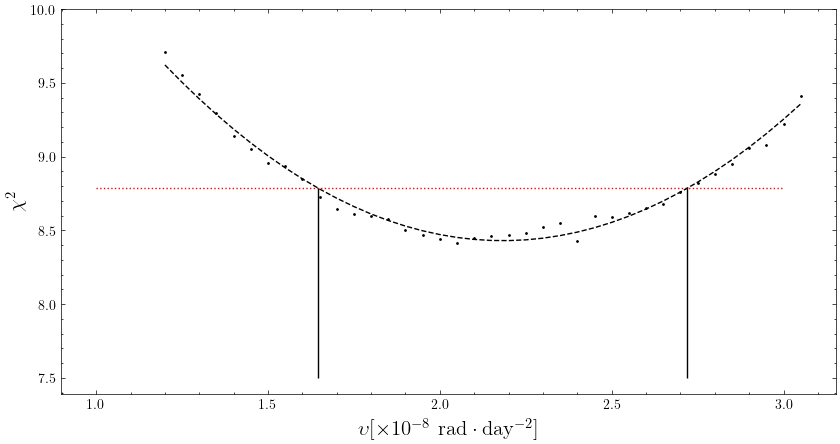}
    \caption{Dependence of the goodness of the fit measured by the $\chi^2$ on the YORP parameter $\upsilon$ for asteroid (85989) 1999 JD6. The dashed curve is a quadratic fit of the data points. The horizontal dotted line indicates a 4.4\% increase in the $\chi^2$ , which defines 1$\sigma$ uncertainty interval of $\pm0.5\times10^{-8}\ \rm{rad\cdot d^{-2}}$ given the number of degrees of freedom.}
    \label{fig10}
\end{figure}

To have an independent estimate, the uncertainty of the $\upsilon$ parameter is also estimated by varying it around its best value and seeing the increase in $\chi^2$. The same method is also used to compute $\chi^{2}$ for different fixed values of $\upsilon$ as shown in Fig.\ref{fig10} for $\upsilon$ between $1.0\times10^{-8}$ and $3.0\times10^{-8}\ \rm{rad\cdot d^{-2}}$. These intervals are larger than those determined by the MC method.

This paper largely eliminates any possible observational errors by weighting the light-curve data and gives a YORP rotational acceleration. In upcoming apparitions, more photometric observations will be required to ensure a robust detection and decrease the uncertainty of the $\upsilon$ value.

\section{Discussion}
\label{sect:discussion}

YORP effect is weak in magnitude, and the short-time-scale observation effect is inconspicuous, making direct detection of YORP effect challenging. Using light-curve data from a sufficiently extended observation span.

In the article, for (1685) Toro we used archived light curves from 1972 to 2021. From this data set, $\upsilon=3.2\times10^{-9}\ \rm{rad/d^2}$ is updated in a YORP value. The formal phase shift in $\Delta T\simeq49$ years produced by the YORP term is $\upsilon(\Delta T)^2/2\simeq29^\circ$. For this value of $\upsilon$, the $\chi^2$ drops by 11\% with respect to $\chi^2$ for $\upsilon=0$. The phase offset is $14^\circ$ when the light curves in 1972 is omitted, and the difference between the YORP model and a constant period model with $\upsilon=0$ is only 3\% in $\chi^2$. A constant period model is well fitted the light-curve data, and no YORP signal is detected, except for the light curves in 1972\citep{dunlap1973} and 1988\citep{hoffmann1990} --- only using the light-curve data from 1996 to 2021. This indicates that the YORP model's direct detection of the YORP effect is strongly dependent on the observation span. 

For (85989) 1999 JD6, the YORP model measured a YORP rotational acceleration of $(2.4\pm0.3)\times10^{-8}\ \rm{rad\cdot d^{-2}}$ only with the optical light curves. If the four light curves from 2000 are exclude, the observations span only 6 years, and the difference between the constant period and YORP model is not significant:the two models fit the data essentially the same, and the phase offset between the models for $2.4\times10^{-8}\ \rm{rad\cdot d^{-2}}$ is only $\sim\ 1^{\circ}$. However, there is also no reasonable reason to suppose the four light curves from 2000 were all time-shifted in the same way to mimic the YORP effect. Additional optical and radar data are required to further confirm this tentative YORP rotational acceleration.

(85989) 1999 JD6 with a tentative YORP detection marks the eleventh direct detection of YORP to date (all detections are reported in Table \ref{tab:Table 3}), and 1999 JD6 is the third contact-binary asteroid with YORP detections (besides Itokawa and 2001 KZ66). 

\begin{table*}
    \centering
	\caption{Summary of all detections of YORP to date. Parameters listed in table: the YORP parameter $\upsilon$, sidereal period (with uncertainty given in parenthesis), secular rotation-period change($dP/dt$), pole orientation($\lambda,\beta$), orbital obliquity, the total observation span(Obs. yrs), and the diameters, whose value was taken from \citet{vishnu2012} for 1999 JD6 .}
	\label{tab:optical light-curves}
	\resizebox{\textwidth}{!}{
	\begin{tabular}{lcccccccl} 
		\hline
		Asteroid& $\upsilon$ & Peroid &$dP/dt$& Pole & Obliquity &Diameters&Obs. yrs& Reference  \\
		     &[$\times10^{-8}\ \rm{rad\cdot d^{-2}}$]&[h]&[$\rm{ms\cdot yr^{-1}}$]&[$^\circ$]&[$^{\circ}$]&[km]&&    \\
		\hline
		YORP&$349\pm30$&0.20283333(1)&-1.25&(180, -85)&174.3&0.113&2001-2005&\citet{lowry2007} \\
		&&&&&&&&\citet{taylor2007} \\
		Eger&$1.1\pm0.5$&5.710156(7)&-3.1&(226,-70)&155.6&1.5&1987-2016&\citet{durech2018} \\
		Apollo&$5.3\pm1.3$&3.065447(3)&-4.3&(50,-71)&161.6&1.4&1980-2005&\citet{kaasa2007} \\
		           &$5.5\pm1.2$&3.065448(3)&-4.5&(48,-72)&162.3&1.45&1980-2007&\citet{durech2008a}\\
		Cacus&$1.9\pm0.3$&3.755067(2)&-2.3&(254, -62)&143.2&1.0&1978-2016&\citet{durech2018} \\
		Geographos&$1.15\pm0.15$&5.223336(2)&-2.7&(58,-49)&149.9&2.56&1969-2008&\citet{durech2008b}\\
		Itokawa&$3.54\pm0.38$&12.132371(6)&-45.4&(128.5,-89.7)&178.4&0.33&2000-2007&\citet{lowry2014} \\
	    Bennu &$264\pm105$&4.2960477(19)&-42.5&(87,-65)&161.0& 0.492& 1999-2012 &\citet{nolan2019} \\
	               &$363\pm52$&4.296007(2)&-58.4&&&&1999-2018&\citet{hergen2019}\\
		2001 KZ66&$8.43\pm0.69$&4.985997(42)&-18.3&(170,-85)&158.5&0.797&2010-2019&\citet{zegmott2021} \\
		Toro&$0.33\pm0.03$&10.197826(2)&-3.0&(75,-69)&160.1&3.5&1972-2021&\citet{durech2022} \\
		        &$0.32\pm0.03$&10.197827(2)&-2.9&(60,-72)&166.7&   -  &1972-2021& This work \\
		1992 SK&$8.3\pm0.6$&7.320232(10)&-38.8&(94,-56)&141.0&8.3&1999-2020&\citet{durech2022} \\
		1999 JD6&$2.4\pm0.3$&7.667749(9)&-12.3&(232, -59)&165.3&1.6&2000-2020& This work \\
 		\hline
	\end{tabular}}
	\label{tab:Table 3}
\end{table*}

The contact-binary asteroids are estimated to make up 15\% to 35\% of the NEA population \citep{benner2006, jacobson2016}. There are three possible formation mechanisms of the contact-binary asteroid 1) Two separate bodies could slowly collided to form a bifurcated shape; 2) The two components collapsed to form one body due to the BYORP effect shrinking their mutual orbit \citep{scheeres2007, jacobson2011}; 3) A rubble-pile asteroid with a weak core undergoes deformation as the rotation rate increases \citep{sanckez2018}. For a small core relative to the asteroid size, and a weak-enough cohesion in the core layer, the deformation is gradually changing with a dent in the surface emerging first. As the asteroid is rotation increases, it stretches and transitions to a shape resembling a contact binary before breaking up into two different sized components. Due to the detection of the YORP effect of 1999 JD6, the former two evolutionary mechanisms seem to contradict this scenario. In addition, the overall surface of 1999 JD6 is uniform with no significant differences in albedo or particle size \citep{kuroda2021}. It further reveals that the two components of the (85989) 1999 JD6 came from one body. To sum up, 1999 JD6 might be a rotationally deformed body currently undergoing the process of rotational fission. The question is whether the evolutionary mechanisms require a fast rotation rate due to the YORP effect, whereas 1999 JD6 has a rotation period close to 8 h. Therefore, it is possible that 1999 JD6’s bifurcated shape was formed during a previous YORP cycle when it had a faster rotation period. The change in the rotational state from a previous YORP cycle to the present is possibly caused by small impacts \citep{Scheeres2018} and small-scale topographical changes \citep{statler2009}. In addition, who the asteroids migrate from obliquity of 0$^\circ$(180$^\circ$) to obliquity of 90$^\circ$ or from 90$^\circ$ to 0$^\circ$(180$^\circ$) \citep{golubov2019}, depends on the shape and rotation state of the asteroids. The current spin-state of 1999 JD6 with a high obliquity of 165.3$^\circ$ is close to an end state of YORP-induced obliquity shift,  but it is more possible that (85989) 1999 JD6 has left the tumbling regime and is now in the process of migrating towards an obliquity of 90$^\circ$.

Under the assumption of zero-conductivity, the shapes of asteroids are roughly classified into four types ($\rm{\uppercase\expandafter{\romannumeral1}}/\rm{\uppercase\expandafter{\romannumeral2}}/\rm{\uppercase\expandafter{\romannumeral3}}/\rm{\uppercase\expandafter{\romannumeral4}}$) according to their response to the YORP rotational moment \citep{vokrou2002}. The behavior of both the spin and obliquity components of YORP for each type of asteroid varies with obliquity.  Among them, only the type $\rm{\uppercase\expandafter{\romannumeral4}}$ asteroids conform to the possible evolution trend of 1999 JD6, that is, the spin and obliquely components of YORP are both positive for obliquities of $\sim$ 150$^\circ$ to 180$^\circ$. 

It is interesting that for three contact-binary asteroids(Itokawa, 2001 KZ66 and 1999 JD6) for which YORP have been detected, a secular rotation-period decrease value is larger than other asteroids(while the YORP strength is not much different). This shows that it is more pronounced to change the rotation rate of irregular or elongated asteroids. Among the eleven asteroids with YORP effect, (54509) YORP has the largest YORP intensity and the fastest rotation period, but its period only decreases by 1.25 $\rm{ms\cdot yr^{-1}}$. It suggests that changing the rotation rate of already fast-spinning asteroids is more complicated.

\section{Conclusion}

In this paper, 8 optical light curves are obtained from the monitoring of (1685) Toro from 11 to 18 February, 2021. With these light-curve data and published optical light curves during the 1972-2016, a robust but very weak YORP signal is got, which is verified based on different photometric datasets. These updated values are as follows:$P=10.197827\pm0.000002 \ \rm{h}, (\lambda,\beta)=(60\pm4^{\circ},-72\pm2^{\circ})$, $\upsilon=(3.2\pm0.3)\times10^{-9}\ \rm{rad\cdot d^{-2}}$ ($1\sigma$ errors).

With published optical light curves, the shape model, spin-state parameters, and a YORP detection for (85989) 1999 JD6 has been derived. These values are as follows: a pole direction in ecliptic coordinates $(232\pm2^\circ, -59\pm1^\circ)$, the sidereal period $P=7.667749\pm0.000009\ \rm{h}$ (for JD 2451728.0), and the YORP rotational acceleration $\upsilon=(2.4\pm0.3)\times10^{-8} \ \rm{rad\cdot d^{-2}}\ (1\sigma \ errors)$. 1999 JD6 is a contact-binary asteroid. It is likely to form the deformation of a rubble-pile asteroid with a weak-tensile-strength core due to YORP spin-up and before that, it might have experienced a fast rotation YORP cycle. We plan to periodically optical monitor 1999 JD6 in the future, and these additional optical observations of 1999 JD6 could be used to refine the YORP detection. A thermophysical analysis is also planned to determine the theoretical YORP strength, which could lead to determination of the density in-homogeneity for 1999 JD6.

As mentioned above, all these asteroids are with a positive YORP value. While these asteroids present large light-curve amplitudes and limit the morphology and observation geometry of asteroids probed, it is dangerous to draw far-reaching conclusions from a limited sample of only eleven objects. Therefore, in the future, it is necessary to continue to enlarge the sample of YORP detection, and 
begin to periodically monitor photometric observations on YORP candidates such as 2100 Ra-Shalom \citep{durech2018}. The main targets include NEAs with a larger period (6$\sim$9 h) and inner main-belt asteroids (MBAs). It could be crucial for constraining or reacquainting theoretical concepts of YORP in planetary science.

\begin{acknowledgements}

This work has been supported by the B-type Strategic Priority Program of the Chinese Academy of Sciences (Grant No.XDB41010104), the National Natural Science Foundation of China (Grant No. 11633009), the Space Debris and Near-Earth Asteroid Defense Research Project (Grant Nos. KJSP2020020204, KJSP2020020102), the Civil Aerospace Pre-research Project (Grant Nos. D020304, D020302) and Minor Planet Foundation.

We thank Professor Josef \v{D}urech for providing the convexinv\_YORP software package and helpful discussion. We also thank Professor PING Yiding for providing observation data of the 0.8-m telescope, and thank Dr. LI Xinran and Dr. CHEN Yuanyuan for their helpful comments to improve the manuscript. This work makes use of the NASA/JPL HORIZONS ephemeris-generating program. All image reduction and processing are performed using the Image Reduction and Analysis Facility (IRAF) \citep{tody1993}. The work makes use of the Asteroid Lightcurve Photometry Database repository \citep{warnersh2011} and the Database of Asteroid Models from Inversion Techniques \citep{durech20}.

\end{acknowledgements}

\section*{Data Availability}
The light curves and python programs utilized in this article are available on CNEOST at \url{http://157.0.0.68:32280/alc/}, 
where they can be found and downloaded by searching for the asteroid by name or number. 
The convex shape model package (convexinv) is available on DAMIT at \url{https://astro.troja.mff.cuni.cz/projects/damit/}. 

\bibliographystyle{raa}
\bibliography{bibtex}

\appendix                  

\section{additional figures}

\begin{figure}
\includegraphics[width=\textwidth]{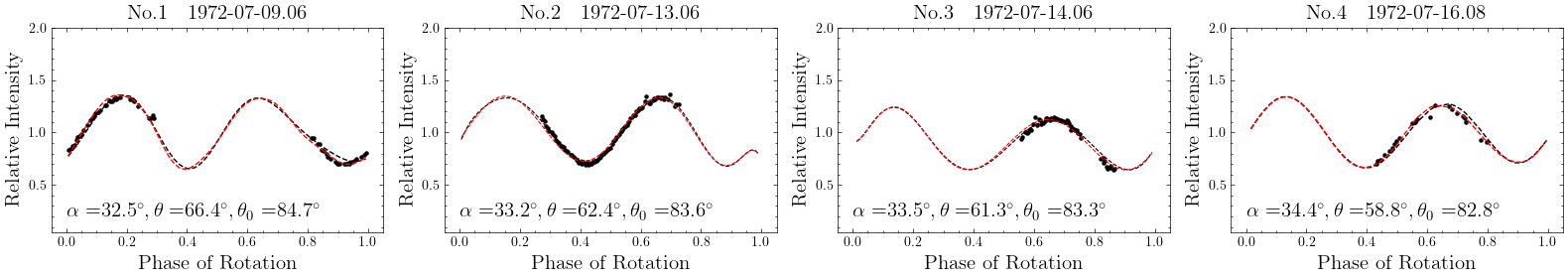}
\includegraphics[width=\textwidth]{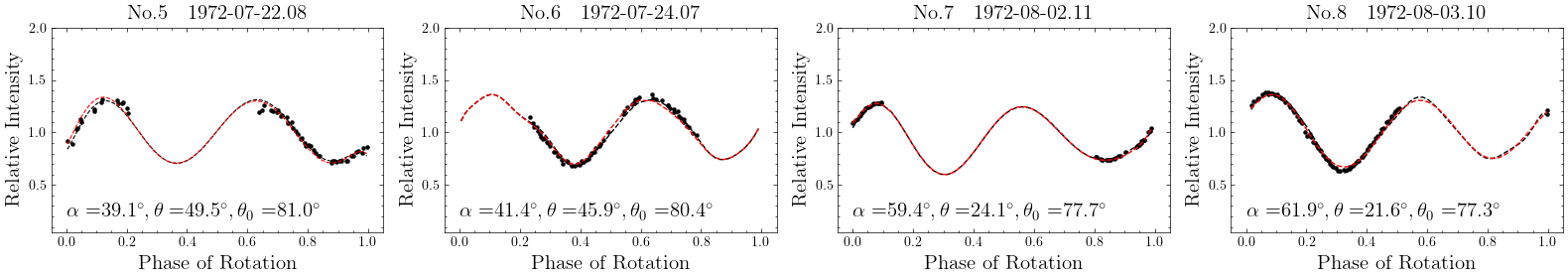}
\includegraphics[width=\textwidth]{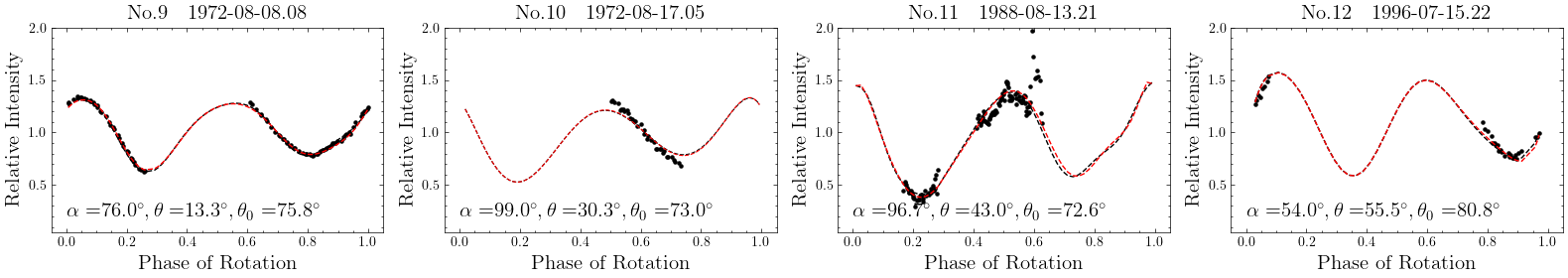}
\includegraphics[width=\textwidth]{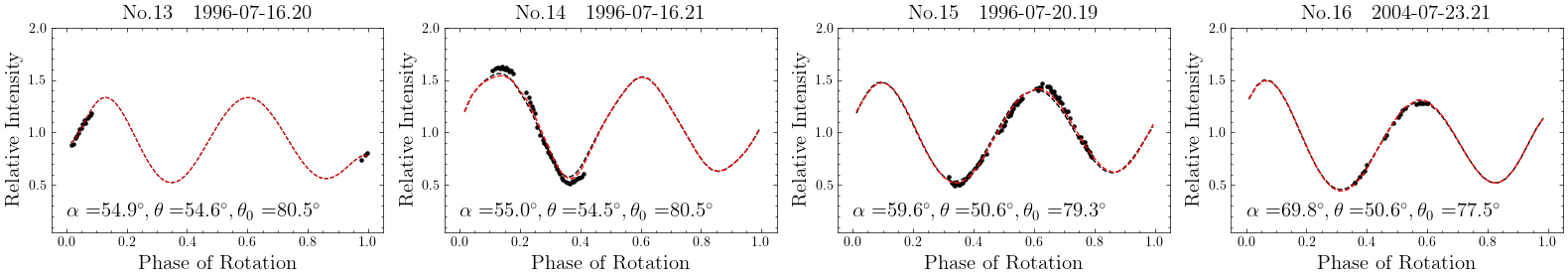}
\includegraphics[width=\textwidth]{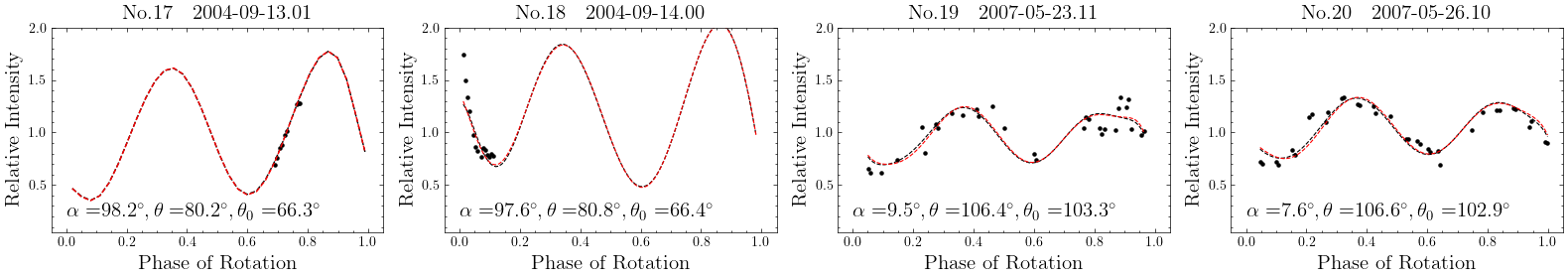}
\includegraphics[width=\textwidth]{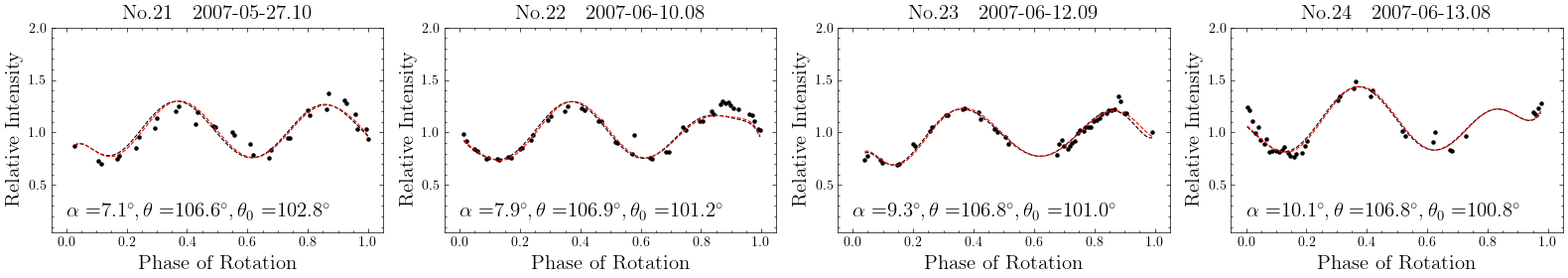}
\includegraphics[width=\textwidth]{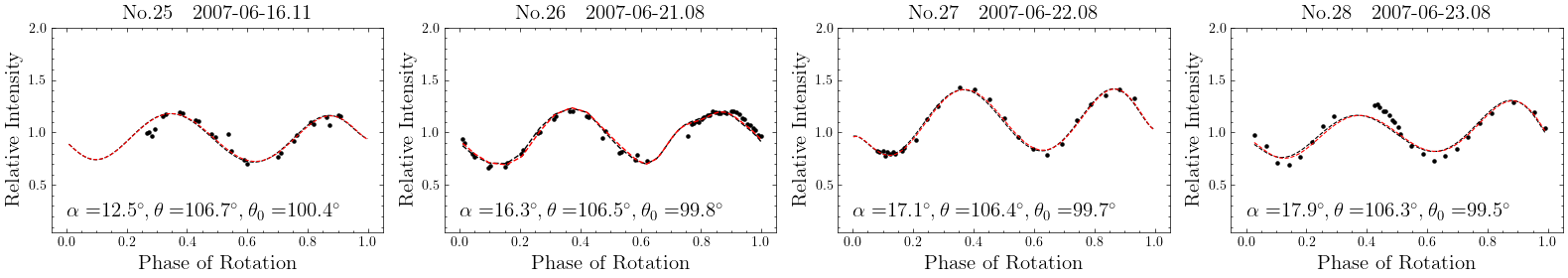}
\includegraphics[width=\textwidth]{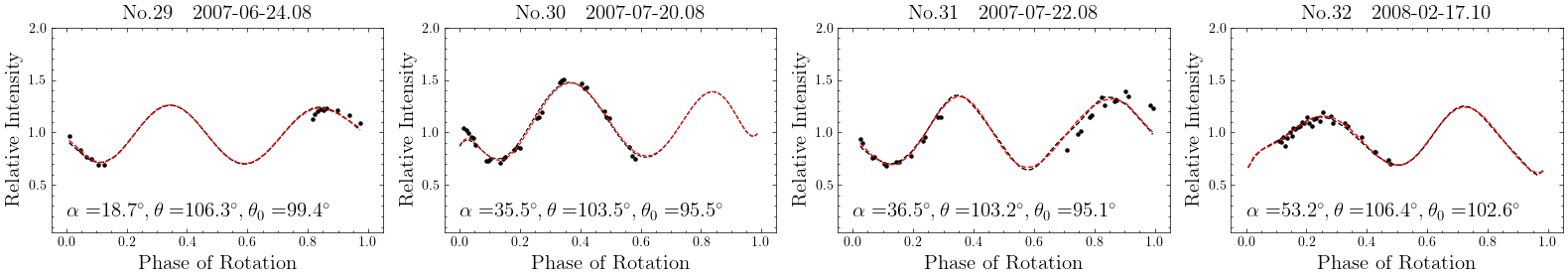}
\caption{All light curves generated with the constant-period of convex-inversion shape model (dotted black curves) and the YORP model (red dashed curves) of asteroid (1685) Toro  plotted over all available light-curve data (black dots).}
\end{figure}
\addtocounter{figure}{-1}
\begin{figure}

\includegraphics[width=\textwidth]{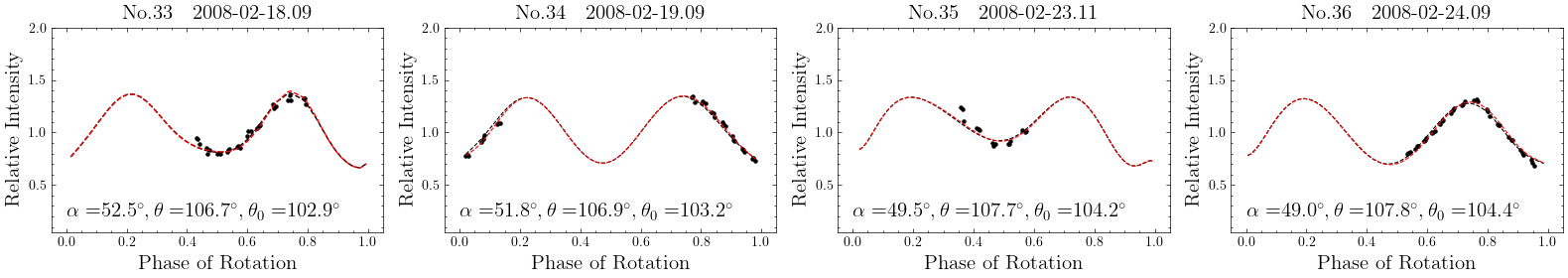}
\includegraphics[width=\textwidth]{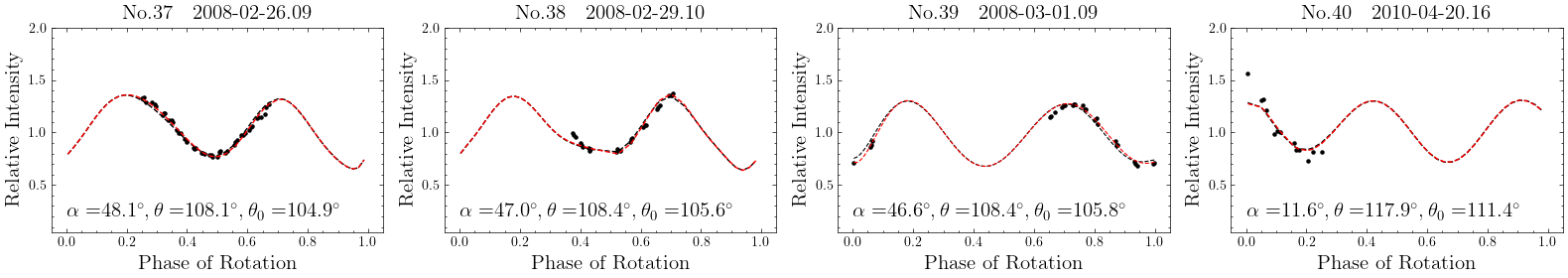}
\includegraphics[width=\textwidth]{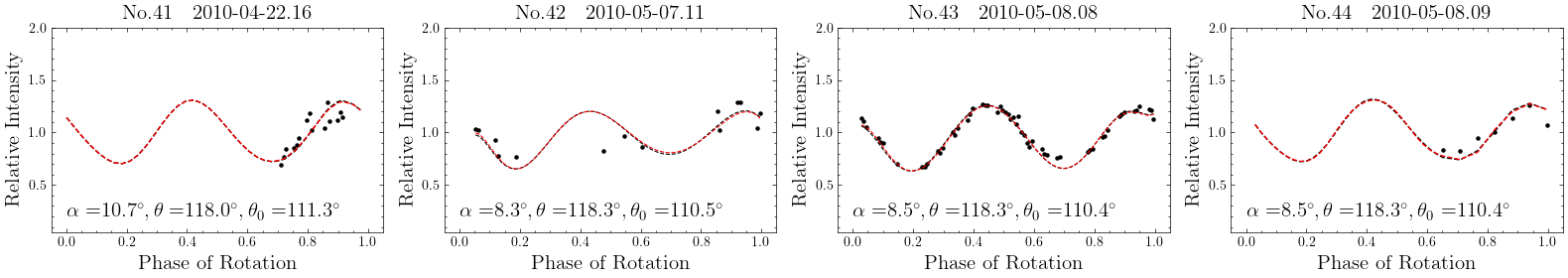}
\includegraphics[width=\textwidth]{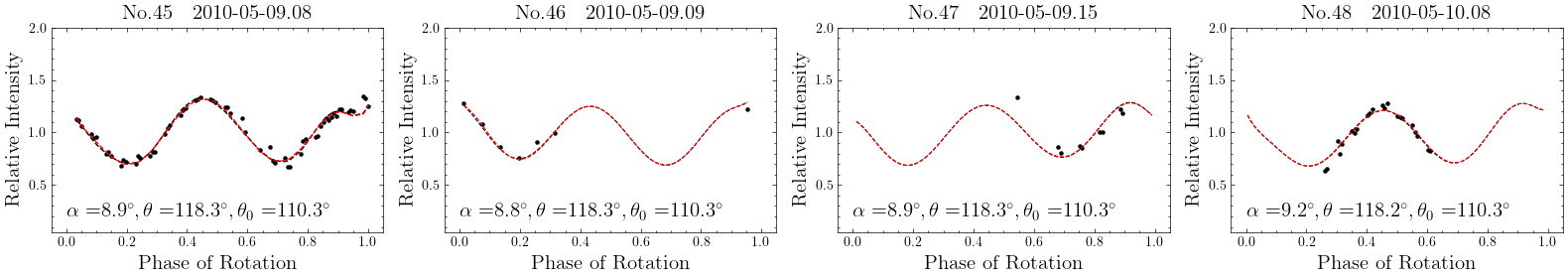}
\includegraphics[width=\textwidth]{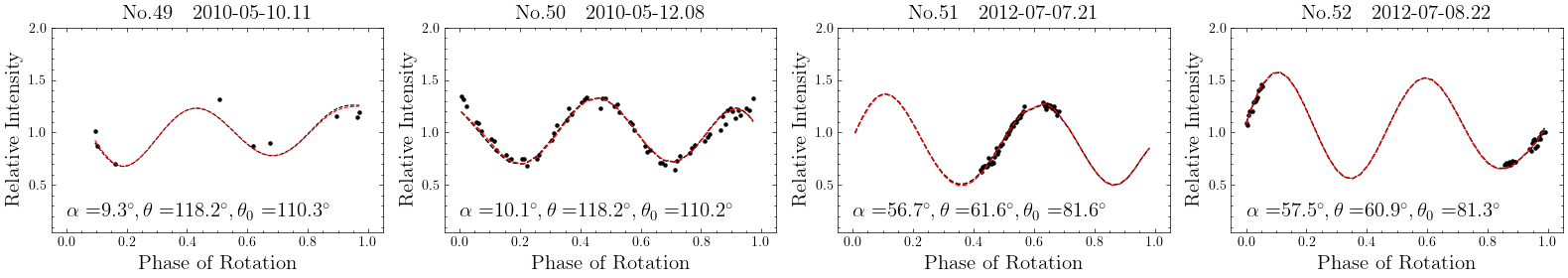}
\includegraphics[width=\textwidth]{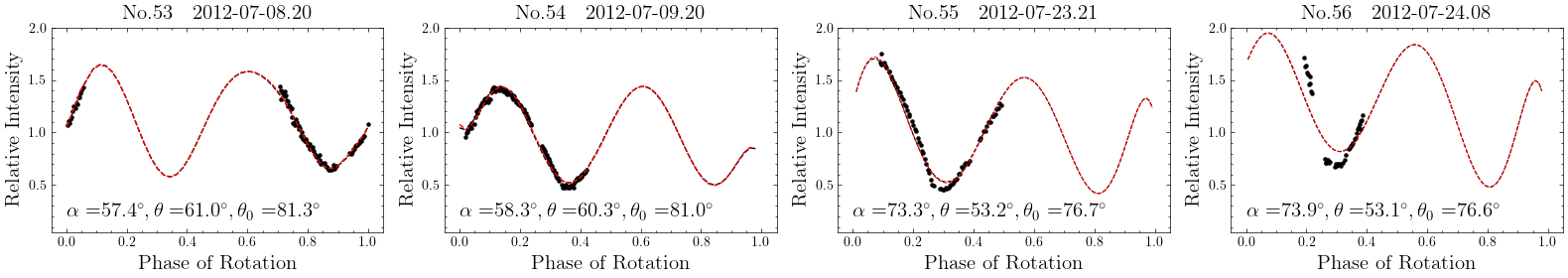}
\includegraphics[width=\textwidth]{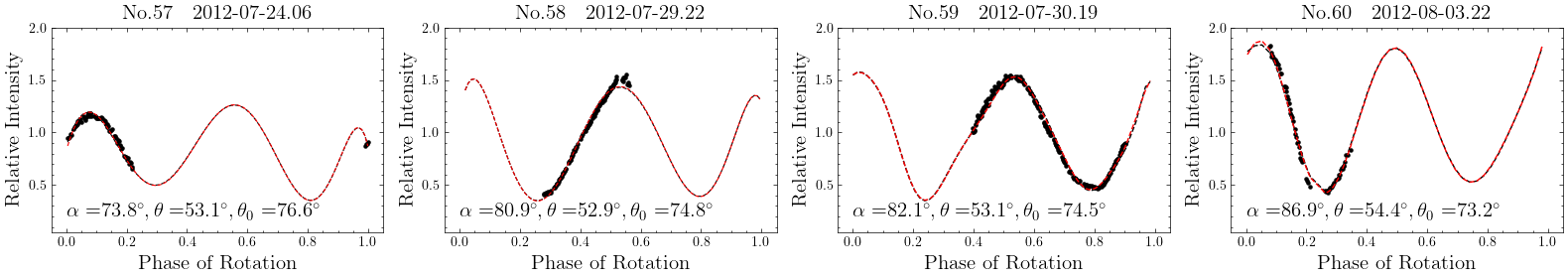}
\includegraphics[width=\textwidth]{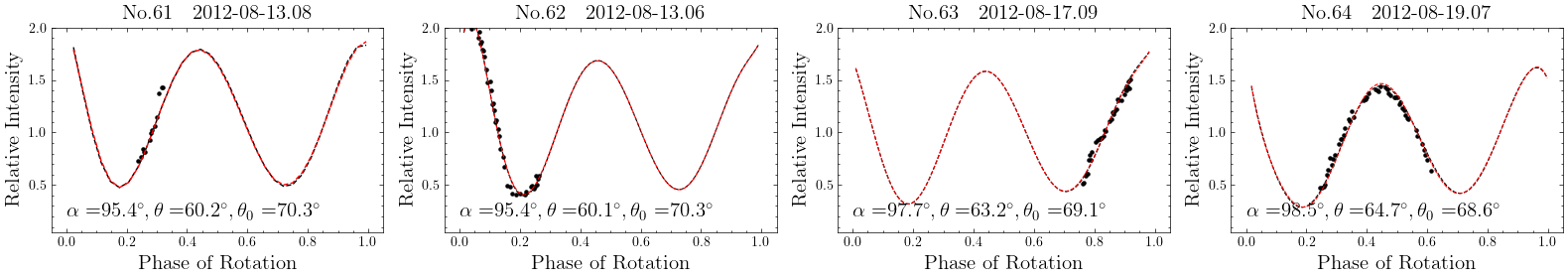}
    \caption{-continued. }
\end{figure}
\addtocounter{figure}{-1}
\begin{figure}

\includegraphics[width=\textwidth]{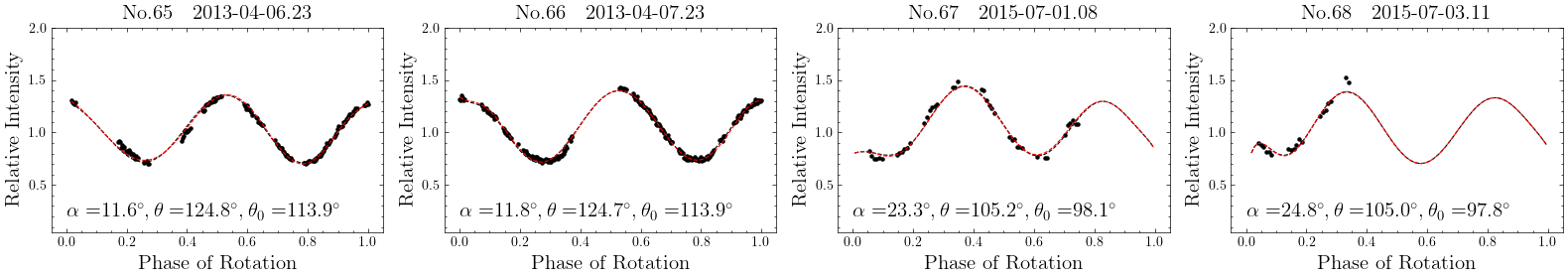}
\includegraphics[width=\textwidth]{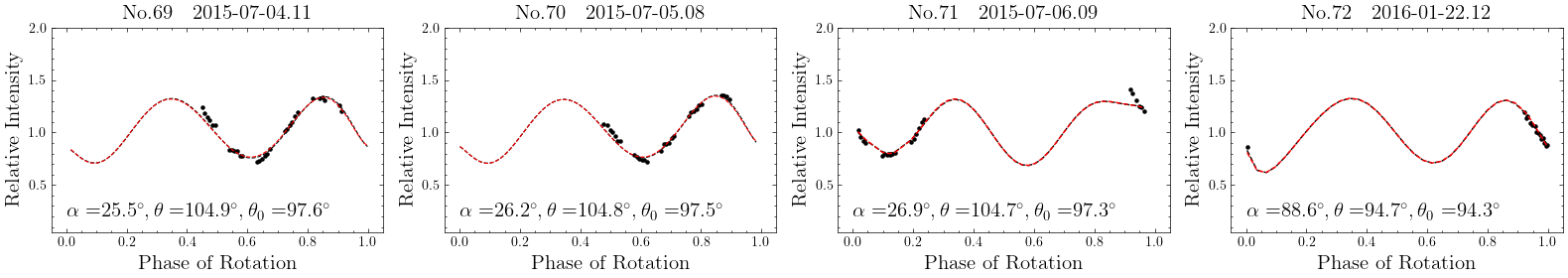}
\includegraphics[width=\textwidth]{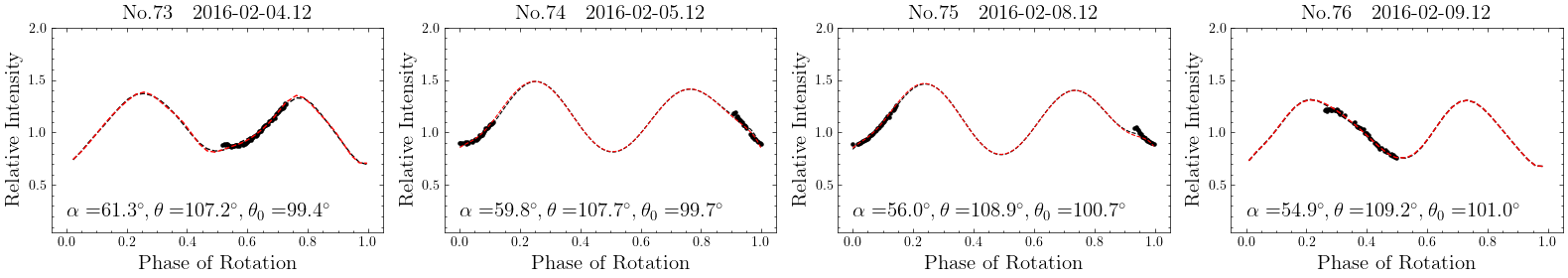}
\includegraphics[width=\textwidth]{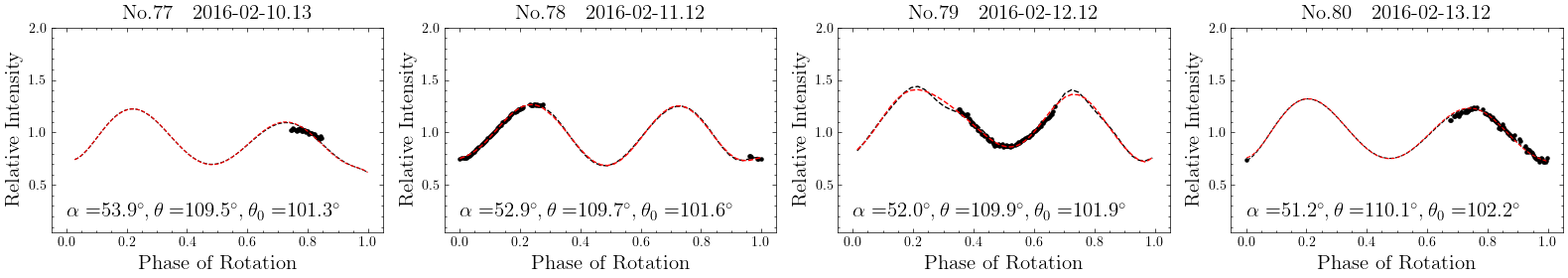}
\includegraphics[width=\textwidth]{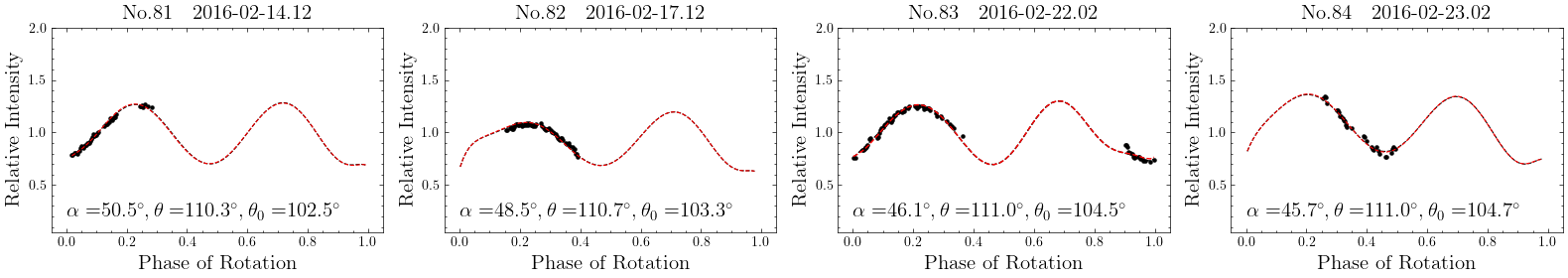}
\includegraphics[width=\textwidth]{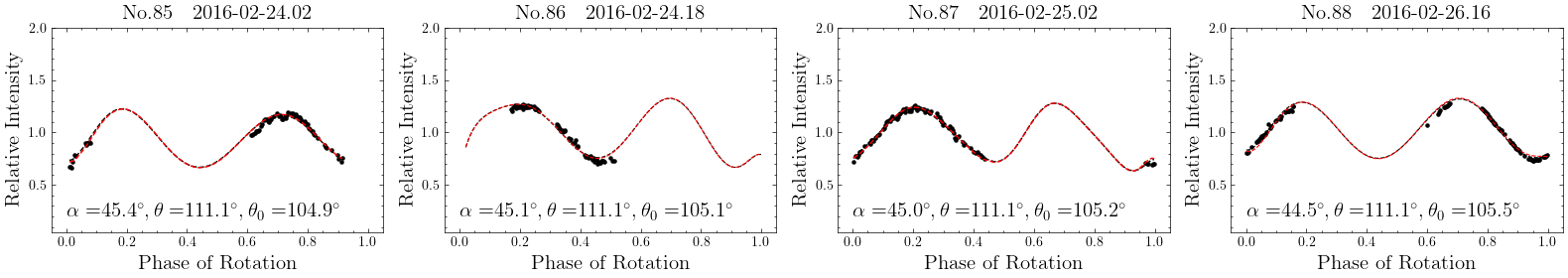}
\includegraphics[width=\textwidth]{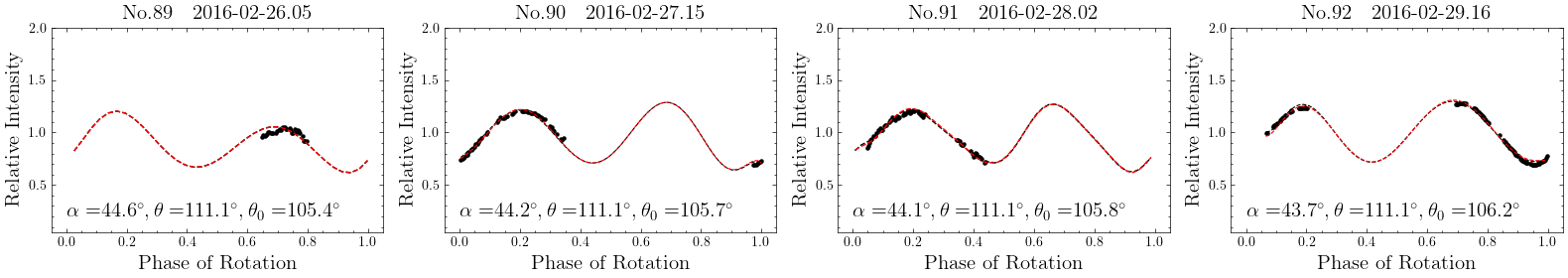}
\includegraphics[width=\textwidth]{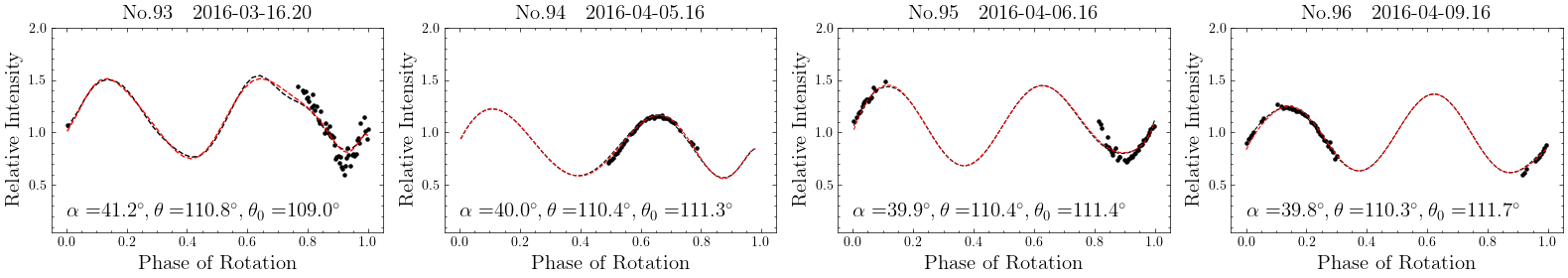}
    \caption{-continued.}
\end{figure}
\addtocounter{figure}{-1}
\begin{figure}
\includegraphics[width=\textwidth]{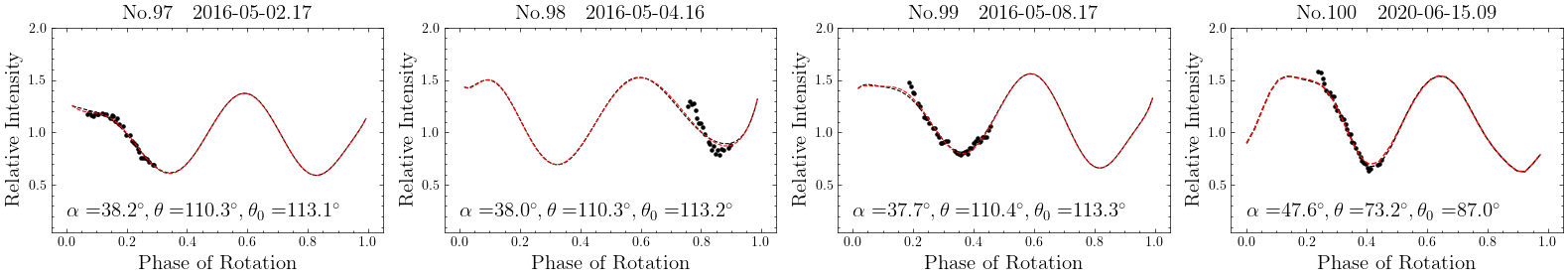}
\includegraphics[width=\textwidth]{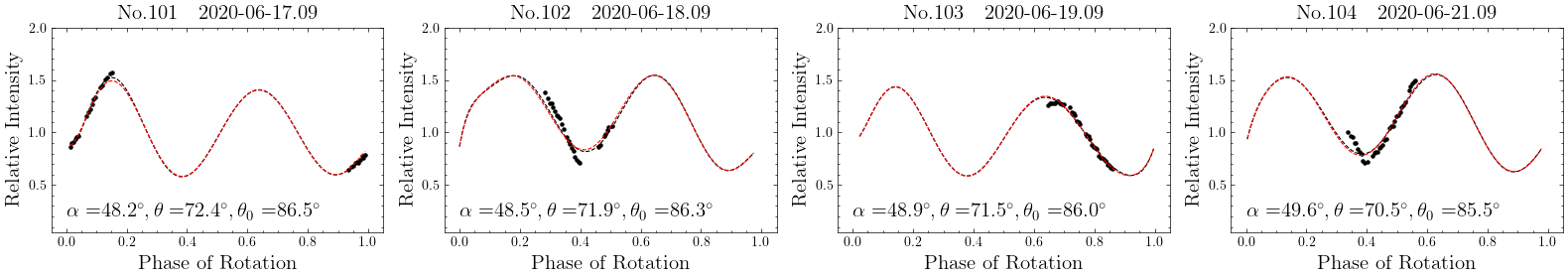}
\includegraphics[width=\textwidth]{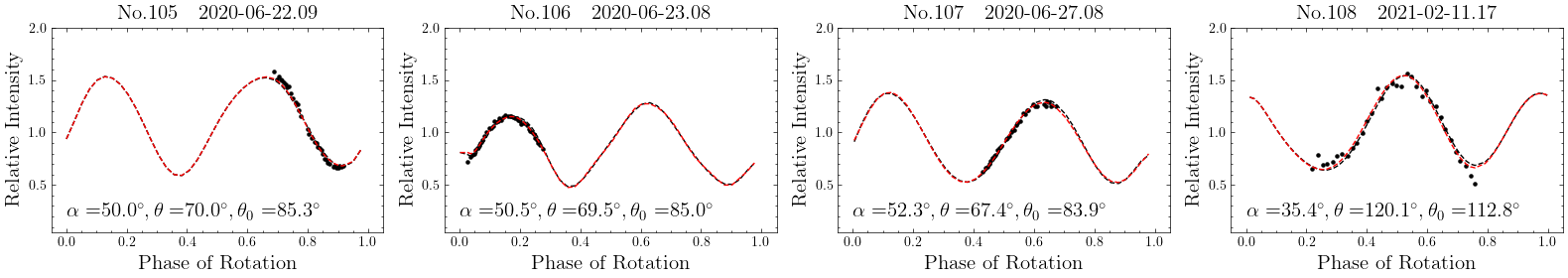}
\includegraphics[width=\textwidth]{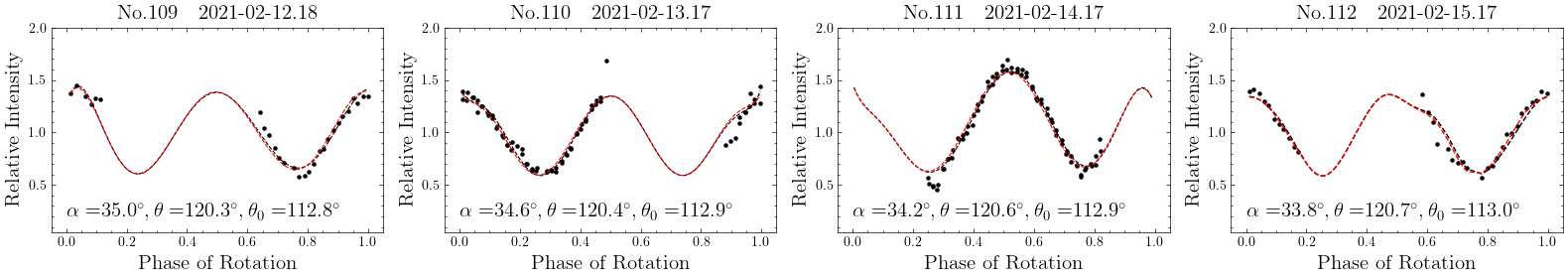}
\includegraphics[width=\textwidth]{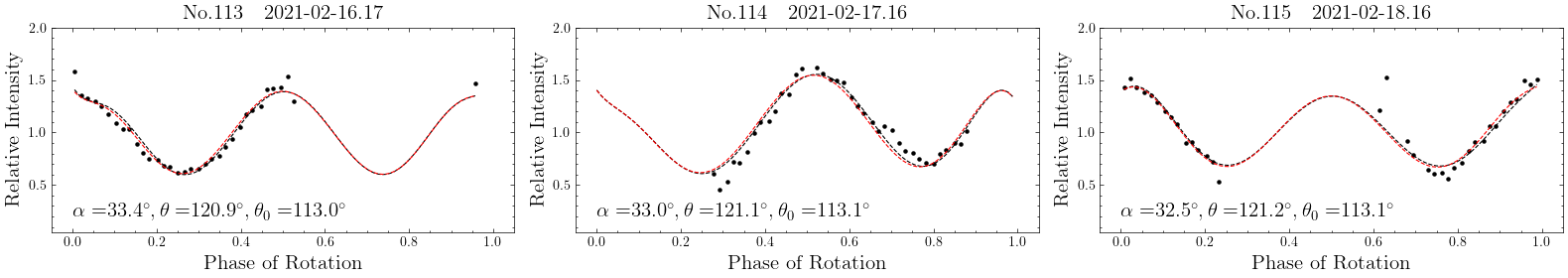}
    \caption{ -continued.}
    \label{fig:Fig.A1}
\end{figure}

\begin{figure}
\includegraphics[width=\textwidth]{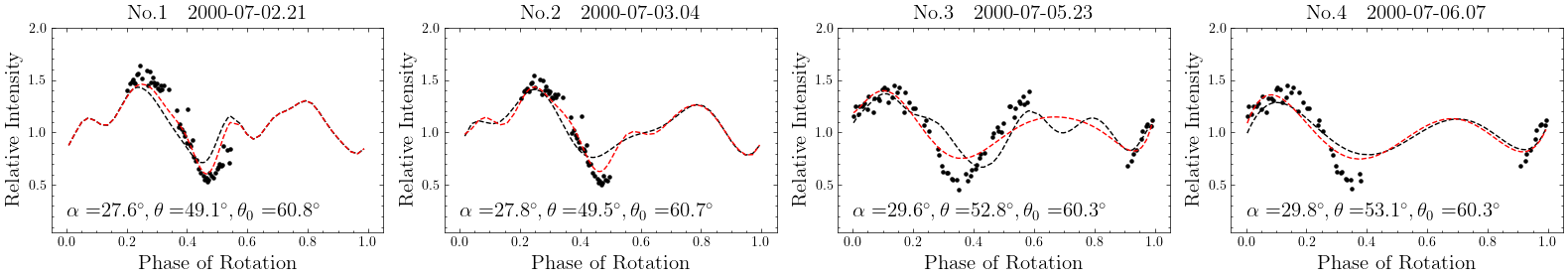}
\includegraphics[width=\textwidth]{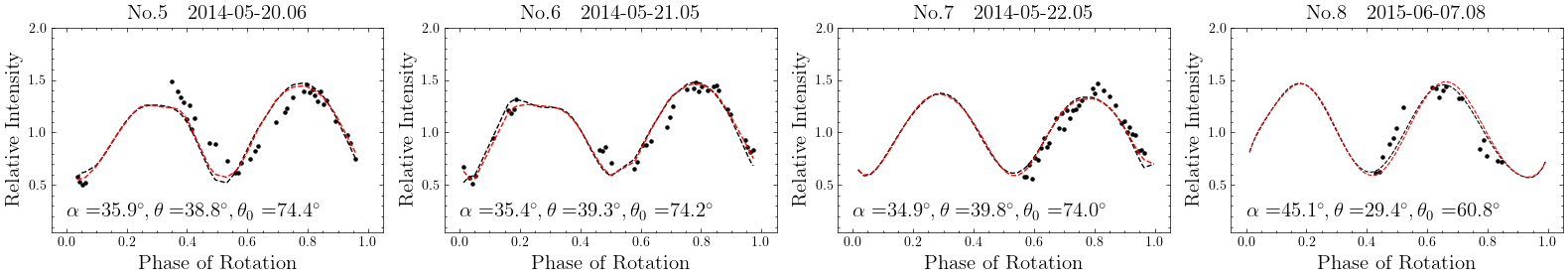}
\includegraphics[width=\textwidth]{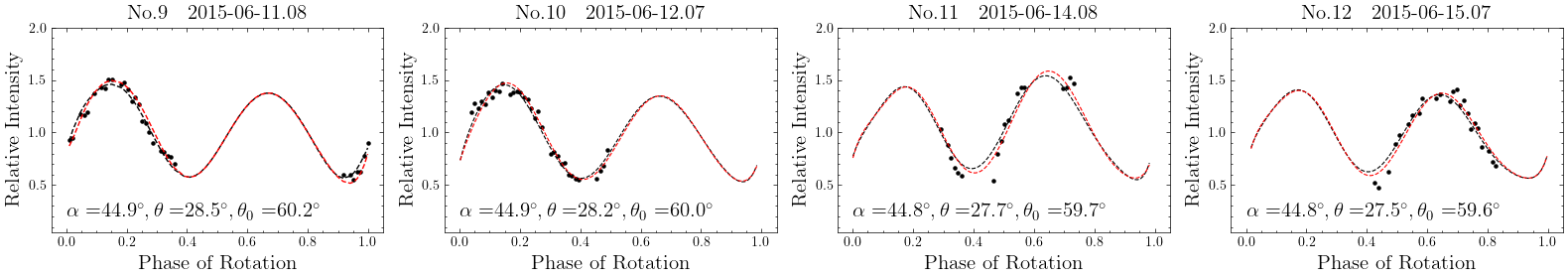}
\includegraphics[width=\textwidth]{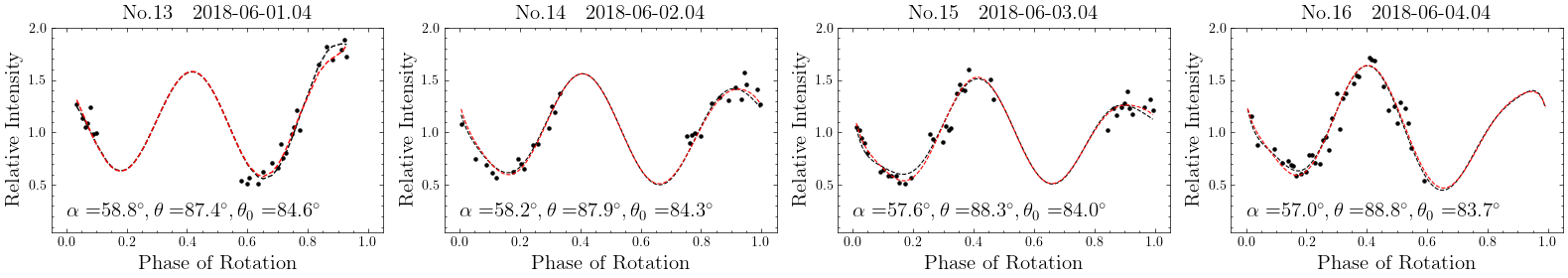}
\includegraphics[width=\textwidth]{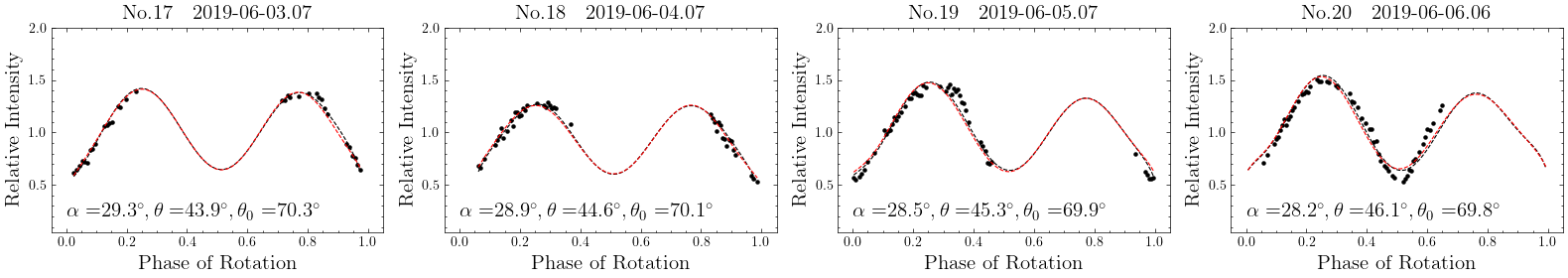}
\includegraphics[width=\textwidth]{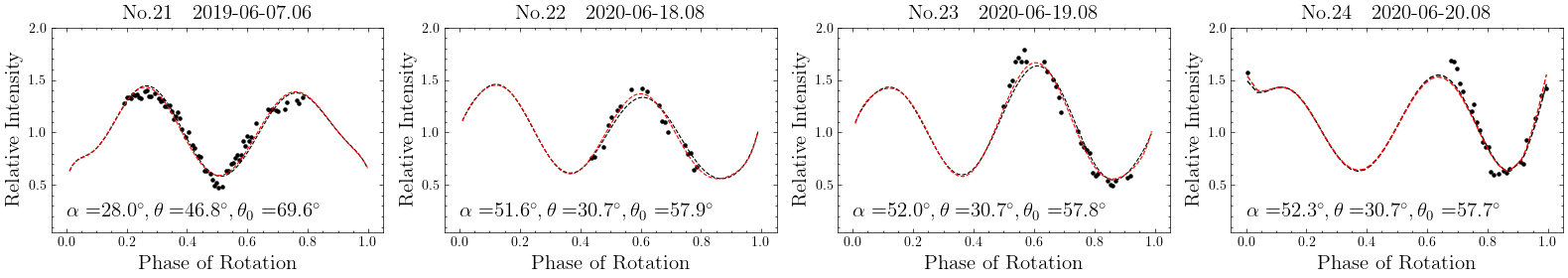}
\includegraphics[width=\textwidth]{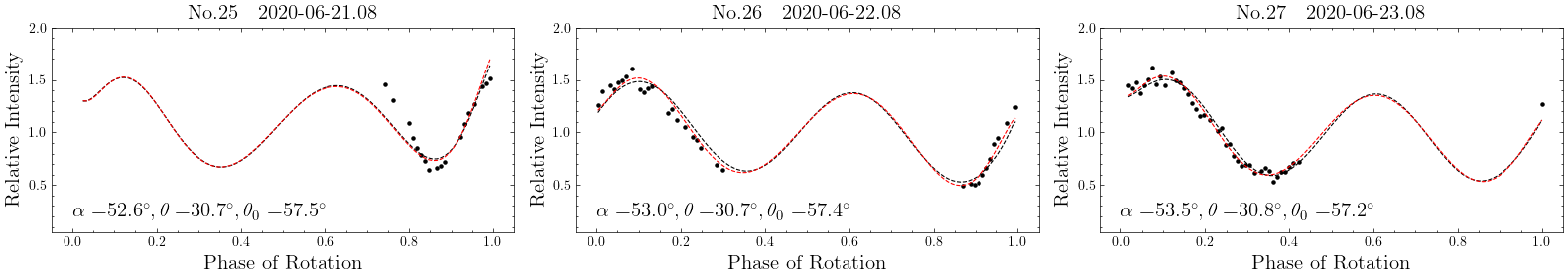}
    \caption{ All light curves generated with the constant-period of convex-inversion shape model (red dots) and the YORP model (blue dots) of asteroid (85989) 1999 JD6  plotted over all available light-curve data (black dots). Light-curve details can be found in Table \ref{tab:Table 2}.}
    \label{fig:Fig.A2}
\end{figure}

\end{document}